\documentstyle[aps,pre,epsf,multicol]{revtex}

\begin{document}

\draft

\title{Phase Bubbles and Spatiotemporal Chaos in Granular Patterns}
\author{Sung Joon Moon$^{1}$,~\cite{email} M. D. Shattuck$^{1}$,~\cite{newaddress}
C. Bizon$^{2}$, Daniel I. Goldman$^{1}$, J. B. Swift$^{1}$,
and Harry L. Swinney$^{1}$~\cite{hlsemail}}
\address{$^{1}$ Center for Nonlinear Dynamics and Department of Physics,
                University of Texas, Austin, TX 78712}
\address{$^{2}$ Colorado Research Associates Division,
		Northwest Research Associates, Inc.
                3380 Mitchell Lane, Boulder, CO 80301}
\date{\today}
\maketitle

\begin{abstract}
We use inelastic hard sphere molecular dynamics simulations and
laboratory experiments to study patterns in vertically oscillated
granular layers. The simulations and experiments reveal that {\em phase
bubbles} spontaneously nucleate in the patterns when the container
acceleration amplitude exceeds a critical value, about $7g$,
where the pattern is approximately hexagonal, oscillating at one-fourth
the driving frequency ($f/4$).
A phase bubble is a localized region that oscillates with a phase
opposite (differing by $\pi$) to that of the surrounding pattern;
a localized phase shift is often called an ${\em arching}$
in studies of two-dimensional systems.
The simulations show that the formation of phase bubbles is triggered
by undulation at the bottom of the layer on
a large length scale compared to the wavelength of the pattern.
Once formed, a phase bubble shrinks as if it had a surface tension,
and disappears in tens to hundreds of cycles.
We find that there is an oscillatory momentum transfer across a kink,
and this shrinking is caused by a net collisional momentum inward
across the boundary enclosing the bubble.
At increasing acceleration amplitudes, the patterns evolve into randomly
moving labyrinthian kinks (spatiotemporal chaos).
We observe in the simulations that $f/3$ and $f/6$ subharmonic patterns
emerge as primary instabilities, but that they are unstable to the
undulation of the layer.
Our experiments confirm the existence of transient $f/3$ and $f/6$ patterns.
\end{abstract}
\nobreak
\begin{multicols}{2}

\section{introduction}
Spatiotemporal chaos, where the physical variables vary in time and space
in a seemingly random way, may arise when a spatially extended system
is driven far from the primary instability~\cite{cross93}.
In some systems, spatiotemporal chaos is understood as 
a complicated evolution of the amplitude field, and the behavior 
can be described in terms of the dynamics of this field~\cite{amplitudefield}.
In some other systems, the dynamics of defects plays a primary role 
in the transition to spatiotemporal chaos; 
cases include Rayleigh-B\'{e}nard convection~\cite{rbc,sdc},
electrohydrodynamic convection~\cite{liquidcrystal},
chemical patterns~\cite{chemical}, and Faraday instabilities~\cite{faraday1}.
Several mathematical models have been proposed to describe the transition
to spatiotemporal chaos in spatially extended physical systems,
including amplitude chaos~\cite{bretherton},
phase turbulence~\cite{shraiman}, defect-mediated 
turbulence~\cite{coullet89,eckmann91}, and invasive defects~\cite{cross95},
but the understanding is still far from complete.

In this paper, we study patterns around the transition 
from the $f/4$ subharmonic hexagonal pattern to spatiotemporal chaos
in granular layers with large aspect ratio ($L/H \gtrsim 10$,
where $L$ is characteristic horizontal size of the layer,
and $H$ is the depth of the layer).
The layers are subject to a sinusoidal oscillation in the direction of gravity.
The oscillation is characterized by two nondimensional control parameters,
$\Gamma = 4\pi^2f^2A/g$ and $f^* = f\sqrt{H/g}$,
where $A$ is the amplitude of the oscillation, $f = 1/T$ is the frequency
of the oscillation, $T$ is the period of the oscillation,
and $g$ is the acceleration due to gravity.
We also define the nondimensional depth of the layer, $N = H/\sigma$,
where $\sigma$ is the diameter of the particle.
Various subharmonic standing wave patterns have been observed
as a function of $\Gamma$ and $f^*$~\cite{thispattern1};
however, the transition to spatiotemporal chaos
has not previously been investigated.
In this study, we show that the transition to spatiotemporal chaos
in granular patterns is due to the intrinsic dynamics of oscillated
granular layers; a large lengthscale undulation of the layer
and an oscillatory momentum transfer across a kink.

The rest of the paper is organized as follows. Section II presents
the methods in the simulation and the experiment.
Kinks, phase bubbles, and randomly moving labyrinths
are described in Section III.
In Section IV a large lengthscale undulation of the layer and
its relation to the nucleation of a phase bubble is described.
Section V discusses how a phase bubble shrinks and disappears.
In Section VI, the prediction and observation of transient $f/3$ and
$f/6$ patterns are presented, and the paper is concluded in Section VII.

\section{methods}

\subsection{Numerical simulation}
In the absence of well-validated macroscopic governing equations
for vertically oscillated granular layers, current theoretical
investigation proceeds at a more basic level, that of individual particles. 
Bizon {\it et al.}~\cite{bizon} developed an event-driven 
inelastic hard sphere molecular dynamics simulation of this system,
by implementing the collision operators in Ref.~\cite{walton}.
This collision model conserves both linear and angular momentum, but allows
energy to be dissipated through inelastic collisions and surface friction. 
The normal coefficient of restitution $e(v_n)$ depends on 
the magnitude of the normal component of relative colliding velocity
$v_n = ({\bf v}_1 - {\bf v}_2)\cdot \hat{\bf r}_{12}$,
where $\hat{\bf r}_{12} = ({\bf r}_1 - {\bf r}_2)/|{\bf r}_1 - {\bf r}_2|$. 
The accurate form of $e(v_n)$ is not known; it was assumed that
$e(v_n) = 1 - Bv_n^{3/4}$ for $v_n$ less than a crossover
velocity $v_c$, and $e(v_n) = e_o$ otherwise.
The value of $B$ was set to make $e(v_n)$ continuous at $v_n = v_c$.
Here we use $v_c = \sqrt{g\sigma}$ and $e_o = 0.7$.
The simulation results are not sensitive to the form of $e(v_n)$ for $v_n < v_c$.
The tangential impulse is given by a coefficient of friction $\mu$
times the normal impulse, with a cutoff corresponding to the crossover
from a sliding contact to a rolling contact. The crossover ratio of the
relative surface velocity after the collision to that of before the 
collision, $-\beta_c$, was set to $-0.35$ as suggested in Ref.~\cite{walton},
and $\mu$ was set to $0.5$.
The values of these parameters were chosen to fit the wavelength
of the pattern obtained in the experiments with lead particles,
for three different control parameter sets.
These fitting parameters reproduced the observed patterns quantitatively
throughout the control parameter space~\cite{bizon}.
The collisions between the grains and the container were treated 
in the same way as the collisions between grains.
The mass of the container was assumed to be infinitely
large compared to that of the granular layer.

We performed three types of simulations: (1) 2D, (2) quasi-2D, and (3) 3D.
We simulated a 2D or quasi-2D layer as vertical cross-section of a 3D layer
(Figs. 3 and 10).
A quasi-2D layer is a 3D layer whose dimension in one direction is
short enough ($\lesssim 10\sigma$) to be homogeneous in that direction;
these simulations run much faster than a fully 3D simulations
yet yield the same statistical information (Figs. 7, 8, and 11(a)).
We performed simulations of 3D layers of square-shape with horizontal
periodic boundary condition (Fig. 11(b)) and of 3D layers of cylindrical
shape with side wall, when we compare with the experiments (Figs. 4, 9).

\subsection{Experiment}
Experiments were conducted with vertically oscillated layers
of granular material consisting of spherical bronze particles of
mean diameter 165 $\mu m$ (Spherical lead particles of diameter
165 $\mu m$ were used only for Fig. 1).
The nondimensional depth of the layer, $N$, was in the range of 5 to 15,
and the aspect ratio $L/H$ ranged from 40 to 150.
Both circular and rectangular containers with various sizes
were used in the experiments.
The container was mounted on an electro-magnetic shaker, 
and it oscillated sinusoidally in the direction of gravity with
a single frequency, in the range 10 - 150 Hz.
The value of $\Gamma$ varied from 0 to $14$.
The container was evacuated to a pressure of 4 Pa to reduce 
the role of interstitial gas.
The container was encircled by a ring of LEDs,
and the images were taken by a digital camera mounted above the container.
A more detailed description of the experimental apparatus is 
found in Ref.~\cite{thispattern2}.

\section{Patterns around a transition to spatiotemporal chaos}

The phase diagram of the patterns in oscillated granular layers has
been reported previously~\cite{thispattern1,bizon,thispattern2}.  
We present a new phase diagram in Fig. 1 that shows the details above
the $f/4$ subharmonic hexagonal pattern regime,
which is the focus of the current study.

\subsection{Single inelastic ball model and temporal dynamics of the layer}
Much of the dynamics of the patterns in this system can be understood from
the single inelastic ball model~\cite{thispattern2,singleball}.
In this Section we review the results of this model in
the $f/2$ and $f/4$ patterns regime.
The single ball model is a one-dimensional model of the oscillated
granular layer, which approximates the center of mass of the layer
as a completely inelastic ball ($e = 0$) on an oscillating plate.

For $\Gamma > 1$, the magnitude of the acceleration
of the container exceeds $g$ during a fraction of the cycle, so that the layer
loses contact with the container when the plate's acceleration becomes $-g$,
and then the layer makes a free flight until colliding with the container later.
In the $f/2$ square/stripe pattern regime, the flight time of the layer
is a fraction of the oscillation period $T$,
and the layer leaves and hits the container every cycle (Fig. 2(a)).
In this regime, the magnitude of the acceleration of the container
at the collision is
less than $g$ (the ball hits below the dot-dashed line in Fig. 2(a)),
and the layer stays on the container until the acceleration becomes $-g$ again
(the intersection of the dot-dashed line and the trajectory of the plate).
The layer leaves the container at the same phase angle of the oscillation
at every cycle; hence the take-off velocity or the flight time is
single-valued. This regime is called {\it Period 1, n = 1},
which means the period of the trajectory is single-valued ({\it Period 1})
and the ball collides with the plate every cycle ({\it n = 1}).
For $\Gamma \gtrsim~4.0$, the trajectory consists of two different flight times,
and the flight time is still a fraction of the period $T$ (Fig. 2(b),
{\it Period 2, n = 1}); it corresponds to the $f/2$ hexagonal pattern.
In this regime, the magnitude of the container acceleration at collision
is larger than $g$ once every other cycle (the ball hits above the dot-dashed
line in Fig. 2(b)).
At this collision, the layer leaves the container immediately,
and the take-off velocity and flight time are smaller than those of
the other cycles, in which the ball stays on the container until
the acceleration of the container becomes $-g$.

For values of $\Gamma$ above $4.5$, the flight time of the layer
exceeds $T$, and the layer hits the container once every other cycle.
In this regime, the layer can hit the container either on odd cycles or on
even cycles; the trajectory becomes degenerate
(the trajectory in the $f/2$ hexagonal pattern regime is also degenerate,
and a $f/2$ hexagonal pattern can have a phase discontinuity defect
which is different from a kink).
The trajectory of single inelastic ball in the $f/4$ square/stripe pattern
regime ($\Gamma \gtrsim~5.5$) is shown in Fig. 2(c) ({\it Period 1, n = 2}).
When $\Gamma$ is increased to $6.5$, the trajectory of the layer
consists of longer and shorter flight times,
and a bifurcation to the $f/4$ hexagonal pattern occurs
(Fig. 2(d), {\it Period 2, n = 2}).
At $\Gamma \sim~8.0$, another bifurcation to {\it Period 1} and {\it n = 3}
state occurs, which corresponds to a $f/3$ flat pattern;
however, the $f/3$ flat pattern has not been observed before,
and the layer has been known to exhibit spatiotemporal chaos in this regime.
We found transient $f/3$ and $f/6$ patterns in this study (see Sec. VI).

Since the layer collides with the container every other cycle
for $\Gamma > 4.5$, domains $\pi$ out of phase may coexist in the layer.
When these opposite phase domains coexist, there is a phase discontinuity
line defect between the adjacent phase domains,
which we call a kink~\cite{thispattern2} (two sides of a phase discontinuity
defect in the $f/2$ hexagonal pattern regime are not at opposite phases,
and this defect is not a kink).
In the experiments, kinks are created by inhomogeneous initial conditions
or external perturbations such as side wall friction or tilt
of the container; see Sec. IV.
A sequence of an $f/4$ pattern with kinks obtained from a 2D simulation
is shown in Fig. 3, where one domain has fully developed
pattern, and the other is nearly flat.
An animated movie is in
{\it http://chaos.ph.utexas.edu/research/moon}.

\subsection{Phase bubbles and randomly moving labyrinths}
As $\Gamma$ is increased further from the $f/4$ hexagonal pattern regime,
the layer exhibits spatiotemporal chaos, which is not included
in the single ball model.
Snapshots obtained from the experiments and simulations of this regime
are shown in Fig. 4.

As $\Gamma$ is increased above a critical value $\Gamma_{pb}$
in the $f/4$ hexagonal pattern regime, 
a small localized region spontaneously changes its phase angle,
and this region becomes enclosed by a kink.
We call this localized region a phase bubble; see Fig. 4(c) and 4(d).
Phase bubbles suddenly (in one cycle) pop up at random locations,
and they shrink and disappear over several tens to hundreds of cycles.
The nucleation rate increases with $\Gamma$, and
the decay time depends on the control parameters and the initial size.
$\Gamma_{pb}$ varies with the depth of the layer and the material used;
it is $7.5$ in the experiment with bronze particles and $N = 5$,
and is $7.1$ in layers of lead particles with $N = 5$, in 3D.
In the simulations we find that the value of $\Gamma_{pb}$ is a few percent
smaller in 2D or quasi-2D layers than in 3D.
A phase bubble is not just a defect in the $f/4$ hexagonal pattern of this system;
it arises due to an inherent instability of the oscillated granular layer
(see Sec. IV).
Phase bubbles are spontaneously nucleated even without side wall friction
or any other external perturbation, while kinks for
$4.5 < \Gamma < \Gamma_{pb}$ are never created spontaneously
without an external perturbation.

As $\Gamma$ is further increased, the nucleation rate of phase bubbles
increases faster than the decay rate, and the layer eventually exhibits
spatiotemporal chaos in the form of randomly moving labyrinths
(Fig. 4(e) and 4(f)).
Note that there are a few kinks of closed form; {\it i.e.}, phase bubbles.
In this regime, labyrinthian kinks move around in the container in a seemingly
random fashion.
Each phase domain of labyrinths collides with the container every other cycle,
and an $f/4$ subharmonic pattern is superposed on it.
The layer behavior does not show any qualitative difference up to
$\Gamma \sim 14$, the highest value investigated in the experiment.

As a phase bubble shrinks, it becomes more circular,
as though it had a surface tension.
This behavior is more clearly observed with phase bubbles in the $f/2$
flat pattern regime, because there is no superposed pattern.
Phase bubbles are not nucleated spontaneously for $\Gamma < \Gamma_{pb}$,
but we can make them from initial conditions with inhomogeneous phase,
by suddenly decreasing $\Gamma$ from the randomly moving labyrinths
regime to the $f/2$ flat pattern regime.
Such phase bubbles in the $f/2$ flat pattern regime are shown
in Fig. 5. An animated movie is in
{\it http://chaos.ph.utexas.edu/research/moon}.

A kink acts on the pattern as if it were a boundary,
and the pattern tries to rotate perpendicular to the kink.
These ``boundaries'' of the layer have irregular shape;
a well-ordered hexagonal pattern does not form in the phase bubble regime
because phase bubbles break the long range order of the pattern
(Fig. 4(c) and 4(d)).
Pattern selection is also affected by the size of a phase bubble,
or the width of a phase domain in the randomly moving labyrinths regime,
because of the ``boundary condition'' of the pattern imposed by the kink.
When a phase bubble is larger than the wavelength of the hexagonal pattern,
the phase bubble has an $f/4$ hexagonal pattern superposed on it.
Otherwise, an $f/4$ stripe pattern is superposed because it is
the only pattern that can fit in such small domains.

\section{Nucleation of Phase Bubbles}
We find in the simulation that for $\Gamma$ slightly below $\Gamma_{pb}$,
the bottom of the layer exhibits an undulation, of which
length scale is much larger than the wavelength of the pattern.
In this Section, we discuss how the undulation leads to
the nucleation of phase bubbles.

In the $f/4$ hexagonal pattern and the phase bubble regime,
the temporal dynamics of the layer is sensitive to a small change in
the flight time.
This is the case also in the single ball model;
in the $f/4$ hexagonal pattern regime, the ball immediately leaves the container
at every other collision, so that the instantaneous velocity
of the container at this collision becomes the take-off velocity of the ball.
Thus, if the velocity is changed slightly and the flight time
is increased, the take-off velocity at the next flight becomes smaller;
this velocity may be too small
for the ball to fly over the container during the following two cycles,
so that the ball collides with the container in the very next cycle (Fig. 6).
In a real layer, the flight time has some fluctuation due to the undulation;
the mechanism in Fig. 6 will create a phase bubble
when the undulation is big enough (Fig. 7).

The power spectral densities (PSD's) of the top and the bottom of a layer are
obtained from the simulation. Spectra for the values of $\Gamma$ slightly
below $\Gamma_{pb}$ show that another peak with small wave number
appears as $\Gamma$ approaches $\Gamma_{pb}$ (Fig. 8).
For smaller $\Gamma$ ($\Gamma \lesssim 6.5$ for this case),
the shape of the bottom of the layer is slaved
to the top of the layer and has the same wavelength as the top;
PSD's have peaks only at the wavelenght of the pattern and its
subharmonics.

The nucleation rate of phase bubbles monotonically increases
with $\Gamma$; we hypothesize that the amplitude or the growth
rate of the undulation increases with $\Gamma$. Above some value of
$\Gamma$ ($> \Gamma_{pb}$),
the nucleation rate exceeds the decay rate, and phase bubbles or kinks
accumulate in the layer until each phase domain reaches
the shortest length scale that cannot have another kink in it;
{\it i. e.}, this sets the width of the labyrinths.
As a result, the layer is filled with a superposition of phase bubbles
which connect to one another, forming non-closed kinks.
We propose that the randomly moving labyrinth pattern
results from the above mechanism. {\it i.e.},
randomly moving labyrinths may be understood as a ``saturated''
state of phase bubbles.

There are two more mechanisms responsible for the creation
of phase bubbles or kinks:
(1) side wall friction and (2) tilt of the layer.
In an infinitely extended layer without side walls, kinks are formed
only by the undulation of the layer, or nucleation of phase bubbles,
if the layer is perfectly level.
In the simulation, side walls are easily eliminated by imposing
horizontal periodic boundary condition; however, the layer size is
always finite in the experiment,
and the side wall effect cannot be completely eliminated.
In the experiment, the friction due to the side walls disturbs
the oscillatory motion of the layer, which often creates kinks.
This effect becomes more and more important as $\Gamma$ is increased.
There are some phase bubbles in contact with the side wall in Fig. 4(c)
and 4(d), which are kinks created by the side wall friction.
In the randomly moving labyrinths regime, the side wall effect is one of
the major reasons for the decay of $f/3$ or $f/6$ transient patterns
(see Sec. VI).
In addition to the above effect, tilt of the container also creates kinks.
This is the main mechanism of the creation of kinks for
$4.5 < \Gamma < \Gamma_{pb}$ in the experiment.

\section{Dynamics of A Phase Bubble}
Once formed, a phase bubble shrinks as if it had
a surface tension, and then disappears (Fig. 5).
We now discuss why this happens.

When the layer collides with the plate,
a density wave forms and propagates across the boundary
of a phase bubble (a kink). 
This density wave initiates a collisional momentum transfer in the
horizontal as well as in the vertical direction, across a kink (Fig. 9).
A sequence of side views of a kink in a 2D layer is in Fig. 10,
which shows the relation for the density wave to the momentum transfer.
At $t = 0$ in Fig. 10, the left half of the layer collides with
the plate (horizontal bar) and is pushed up.
It becomes compact and nearly static with respect to the plate,
acting as if it were a part of the container.
At this moment, the rest of the layer is still falling and is dilated.
The two parts of the layer interact at the kink,
and a density wave forms at the interface due to the large density gradient.
The density wave propagates toward the dilated part until the compact part
loses contact with the plate. The density wave drives a collisional
momentum transfer and pushes the interface toward the dilated part
(the kink is shifted rightward
at $t = 1.2T$ compared to its earlier location at $t = 0$).
In the next cycle, the same process occurs in the opposite
direction, and the kink is pushed back to its original position
(not shown in Fig. 10), provided that the phase difference is exactly $\pi$
(then both parts of a 2D layer are symmetric).
As a result, the momentum flux across a kink in a 2D or quasi-2D layer oscillates
symmetrically with a period of $2T$, with no net translational motion;
thus a phase bubble in a 2D or quasi-2D layer never shrinks but only
oscillates symmetrically (Fig. 11(a)).

The mechanism of the density wave and the momentum transfer 
across a kink in a 3D layer is the same as in a 2D/quasi-2D layer; however,
in a 3D layer, there is a geometric effect which is absent
in a 2D/quasi-2D layer: the two sides of a kink are not symmetric
unless the kink is straight.
As the particles at the front of the density wave are nearly static
with respect to the plate, the momentum flux is roughly proportional
to the number of particles at the front of the density wave
and the velocity of the container.
For a kink with nonzero local curvature, the number of particles 
at the front of the density wave for the two successive cycles are different,
even if the two sides of the kink are exactly $\pi$ out of phase.
Thus the momentum transfer in one direction is always larger than
that of the other. This leads to an asymmetric momentum transfer,
or an asymmetric oscillatory motion of a kink.
Consequently, the total net momentum flux across the boundary of
a phase bubble over a multiple of $2T$ is always inward,
because its boundary is a kink of a closed shape.
As a result, a phase bubble shrinks and disappears.
An asymmetric oscillation of a boundary of a phase bubble is shown
in Fig. 11(b).
The difference in the numbers of particles at the interface of
the density wave for the two consecutive cycles is approximately
proportional to the local curvature of the kink.
Thus, the speed of a kink is roughly proportional to
the local curvature, leading to an effective surface tension.

There are other mechanisms driving the motion of a kink in the experiment,
which are minor compared to the previously discussed curvature-dependent
geometrical effect: (1) {\it non-$\pi$ phase difference effect} and
(2) {\it the finite mass ratio effect}.
Firstly, we have assumed in the above discussion that both sides of a kink
are exactly $\pi$ out of phase; if this were not the case, the impact from
the container for two successive cycles would be different,
and the oscillation of a kink would be asymmetric,
even if the local curvature were zero.
This is why a secondary forcing can control the motion of a straight kink,
as was experimentally observed by Aranson {\it et al.}~\cite{aranson}.
Secondly, if the difference in the mass of each phase domain
of the layer is not negligible compared to mass of the container,
the impact by the container on both domains of the layer are different,
and a net translational motion of a kink is induced,
even if it has zero local curvature and the phase difference is $\pi$.
We call this {\it the finite mass ratio effect}.
In the experiment, a kink in the $f/2$ flat pattern regime
eventually travels to the center of the container,
until the impacts from the container on both phase domains are balanced;
it was observed in Refs.~\cite{thispattern2,aranson}.
This effect is absent in the simulation, because the mass of the container
is assumed to be infinitely large compared to the mass of the layer.

\section{Transient $\lowercase{f}/3$ and $\lowercase{f}/6$ subharmonic patterns}
Until now, no standing wave pattern has been observed above the $f/4$ hexagonal
pattern regime in the phase diagram in Fig. 1. In this Section
we present the first observation of the $f/3$ flat and $f/6$ square/stripe patterns.
These patterns were found first in the simulations,
which motivated the experiments.
We start the discussion with the single ball model.

The single ball model predicts an infinite cascade of bifurcations with
increasing natural number $n$, for increasing $\Gamma$ (there are small
chaotic windows along $\Gamma$, but this is not important in this discussion).
For the granular layers, these bifurcations correspond to the following
patterns:\\

\begin{picture}(200,100)(0,0)
\put(20,90){$f/n$ flat}
\put(20,50){$f/(2n)$ squares/stripes}
\put(20,10){$f/(2n)$ hexagons}
\put(40,80){\vector(0,-1){15}}
\put(40,40){\vector(0,-1){15}}
\put(105,12){\line(1,0){45}}
\put(150,12){\line(0,1){81}}
\put(150,93){\vector(-1,0){85}}
\put(160,50){$n~\rightarrow~n+1$}
\end{picture}
\\
where the initial $n$ is $1$.
This model predicts a bifurcation from an $f/4$ to an $f/3$ state
({\it Period 1, n = 3}) at around $\Gamma = 8.0$,
however, in the experiments, the cascade of bifurcation stops at $n = 2$.

We discussed how the undulation of the layer leads to the nucleation
of phase bubbles in Sec. IV.
If the undulation of the layer could be avoided,
we expect that $f/3$ flat and $f/6$ square/stripe patterns would exist.
To test this conjecture, we performed a series of numerical simulations
of a small layer ($10\sigma \times 10\sigma$ and $20\sigma \times 20\sigma$
with $N = 6$) in the randomly moving labyrinths regime
for ($\Gamma, f^*$) = (8 - 11, 0.8 - 1.0),
using periodic horizontal boundary conditions.
In these simulations, we observed that the layer collides with
the plate every third cycle, and that the dynamics is stable up to
order of $1000$ cycles;
we conclude that if a larger layer followed this behavior, 
$f/3$ or $f/6$ subharmonic patterns would exist.

In larger layers, the undulation of the layer cannot be completely avoided,
but can be suppressed at least during a short time
by preparing a flat and compact layer as an initial condition;
when $\Gamma$ is quickly increased from the flat pattern
regime to where $f/3$ or $f/6$ subharmonic patterns are predicted by
the single ball model ($8 \lesssim \Gamma \lesssim 11$),
a $f/3$ or $f/6$ pattern is found in the simulation.
Later these transient patterns were found in the laboratory experiments
as well, as Fig. 12 and 13 illustrate.
These patterns emerge as the primary instability of the layer
in this regime, but they are invaded and overtaken by either 
kinks formed due to the side wall friction (Fig. 12)
or by the undulation of the layer (Fig. 13).
As a result, the domain of these transient patterns gradually decreases and 
is eventually taken over by randomly moving labyrinths.
These transient patterns last up to several hundreds of cycles, 
depending on the initial condition, control parameters, and the system size. 

\section{Discussion}
We have shown that phase bubbles play a critical role in the
transition to spatiotemporal chaos in the patterns formed by vertically
oscillated granular layers. Phase bubbles spontaneously form first in the
f/4 hexagonal pattern regime as the acceleration $\Gamma$ is increased. At
larger $\Gamma$, the rate of nucleation of bubbles grows faster than the
decay rate, leading ultimately to a spatiotemporally chaotic pattern of
randomly moving labyrinths.  In a qualitatively similar way, the formation
of defects has been found to lead to spiral defect chaos~\cite{sdc}
and to chaos in a model with invasive defects~\cite{cross95}.

We have investigated the mechanism of the nucleation and the dynamics
of phase bubbles and randomly moving labyrinths,
using inelastic hard sphere molecular dynamics simulations and experiments.
We have found that a vertically oscillated granular layer has
a large scale undulation, even without interstitial gas (Sec. IV).
The undulation constitutes an inherent feature of oscillated
granular layers, like the standing wave pattern formation,
and cannot be avoided;
above some critical value of $\Gamma$ in the $f/4$ hexagonal pattern regime, 
the undulation of the layer gives rise to the nucleation of a phase bubble.
The spontaneous nucleation of kinks in vertically oscillated 2D or quasi-2D
granular layers for large $\Gamma$ has been called
``subharmonic instability''~\cite{douady} or ``arching''~\cite{wassgren,lan},
but the mechanism has not been elucidated.
These two phenomena are all essentially phase bubbles
in 2D or quasi-2D oscillated granular layers.

We also found that a kink of nonzero curvature has a net translational motion
due to the asymmetric collisional momentum transfer across a kink (Sec. V).
The local speed of a kink is roughly proportional to the local curvature;
a phase bubble shrinks as if it had a surface tension
and then disappears, because it is a kink of a closed form.
We showed that shrinking of a phase bubble and translational motion
of a kink are essentially the same phenomena.

Finally, based on the understanding of kinks,
we predicted transient $f/3$ and $f/6$ subharmonic
patterns and observed them for the first time (Sec. VI).

\section*{acknowledgments}

The authors thank Professor W. D. McCormick, E. Rericha, B. Lewis,
J. Bougie, and N. Peffley for helpful discussions.
This work was supported by the Engineering Research Program 
of the Office of Basic Energy Sciences of the Department of Energy
(Grant DE-FG03-93ER14312).

\pagebreak

\begin{figure}
\epsfxsize=.98\columnwidth
\centerline{\epsffile{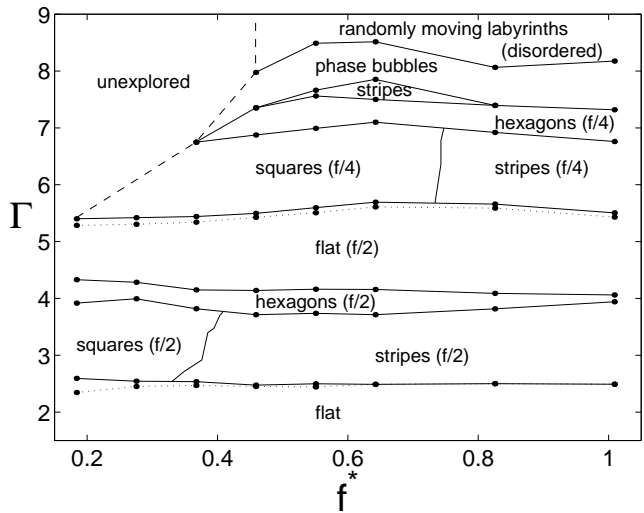}}
\caption{Phase diagram obtained from an experiment 
with bronze particles of diameter $\sigma = 165~\mu m$ and 
nondimensional depth $N = 5$,
in a circular container with diameter $L = 770\sigma$,
showing particularly the details for $\Gamma > 7.0$.
Solid lines denote the transitions for increasing $\Gamma$,
while the dotted lines denote decreasing $\Gamma$.
}
\end{figure}
\pagebreak

\begin{figure}
\epsfxsize=.98\columnwidth
\centerline{\epsffile{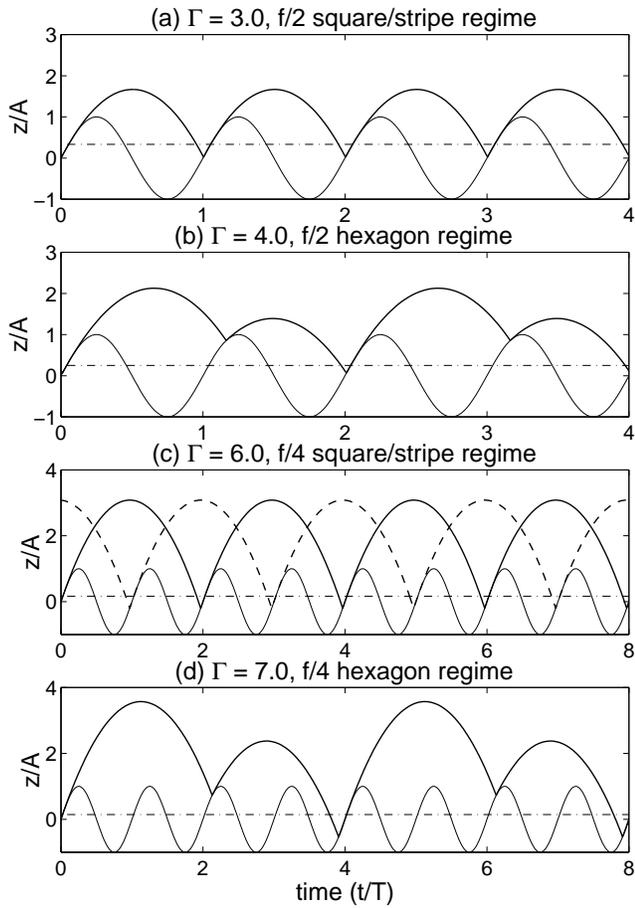}}
\caption{Temporal trajectory of a completely inelastic ball ($e = 0$)
on an oscillating plate, which models the trajectory of
the center of mass of the layer.
The solid sinusoidal curve is the trajectory of the plate.
The ball leaves the plate when the acceleration of the plate
becomes $-g$; {\it i. e.}, where the horizontal dot-dashed line intersects
with the trajectory of the ball.
If the ball collides with the plate above the dot-dashed line
(in (b) and (d)), it leaves the plate immediately.
}
\end{figure}
\pagebreak

\begin{figure}
\epsfxsize=.99\columnwidth
\centerline{\epsffile{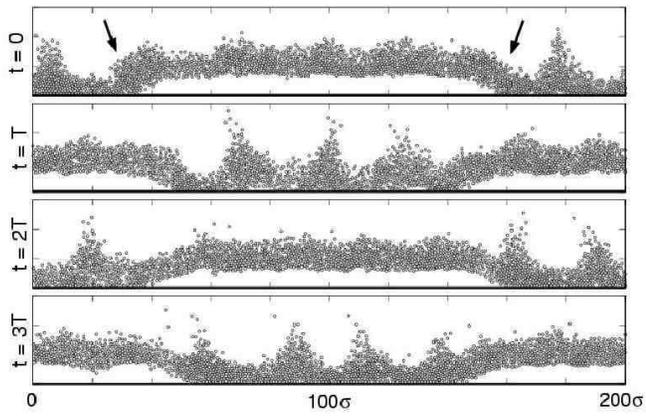}}
\caption{A side view of a 2D layer of an $f/4$ pattern with two kinks,
obtained in the simulation for $\Gamma = 6.5$, $f^* = 0.8$, and $N = 8$.
The middle part and the rest of the layer oscillate $\pi$ out of phase, 
and phase discontinuities between these two phase domains are called
kinks (indicated by arrows).
The container (horizontal bar at the bottom) is at its minimum height
at the phase angle at which these figures are taken.
Horizontal boundary is periodic.
}
\end{figure}
\pagebreak

\begin{figure}
\centerline{{\sf EXPERIMENTS \hskip 1.truein SIMULATIONS}}
\centerline{{\sf f/4 hexagonal pattern}}
\centerline{\epsfxsize=.49\columnwidth \epsffile{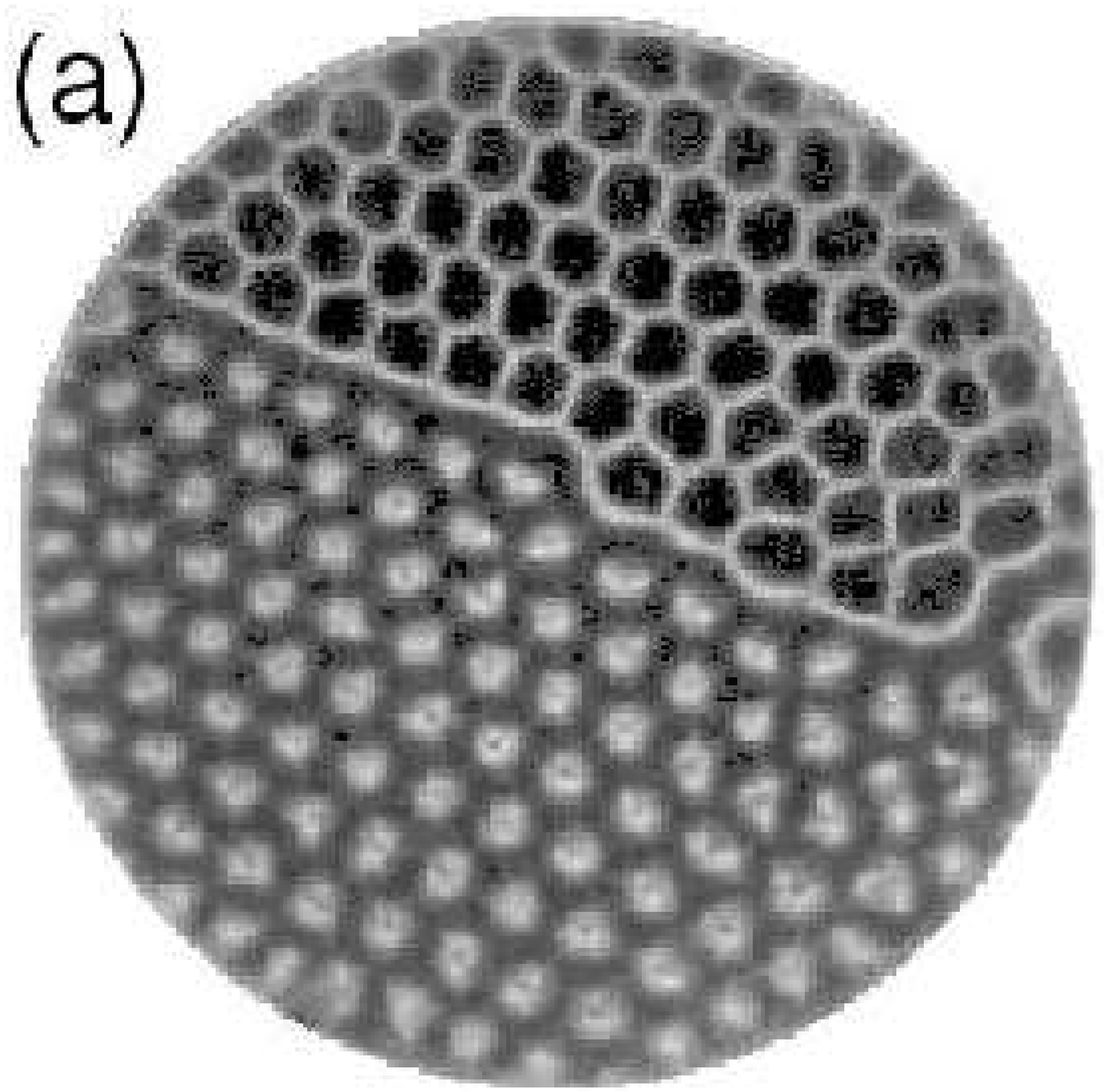}
\hskip .01\columnwidth \epsfxsize=.49\columnwidth \epsffile{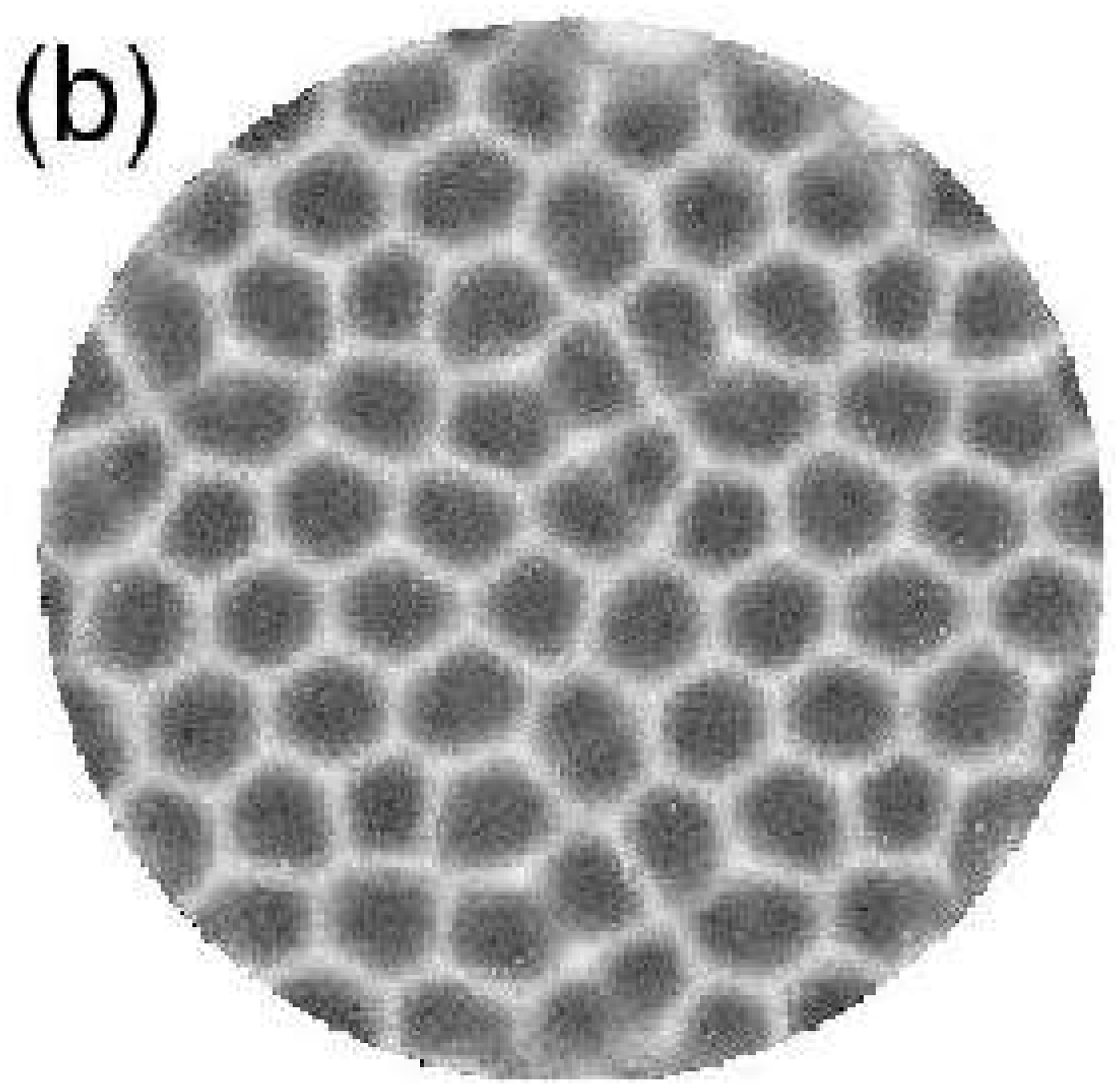}}
\centerline{{\sf Phase bubbles in f/4 hexagonal pattern}}
\centerline{\epsfxsize=.49\columnwidth \epsffile{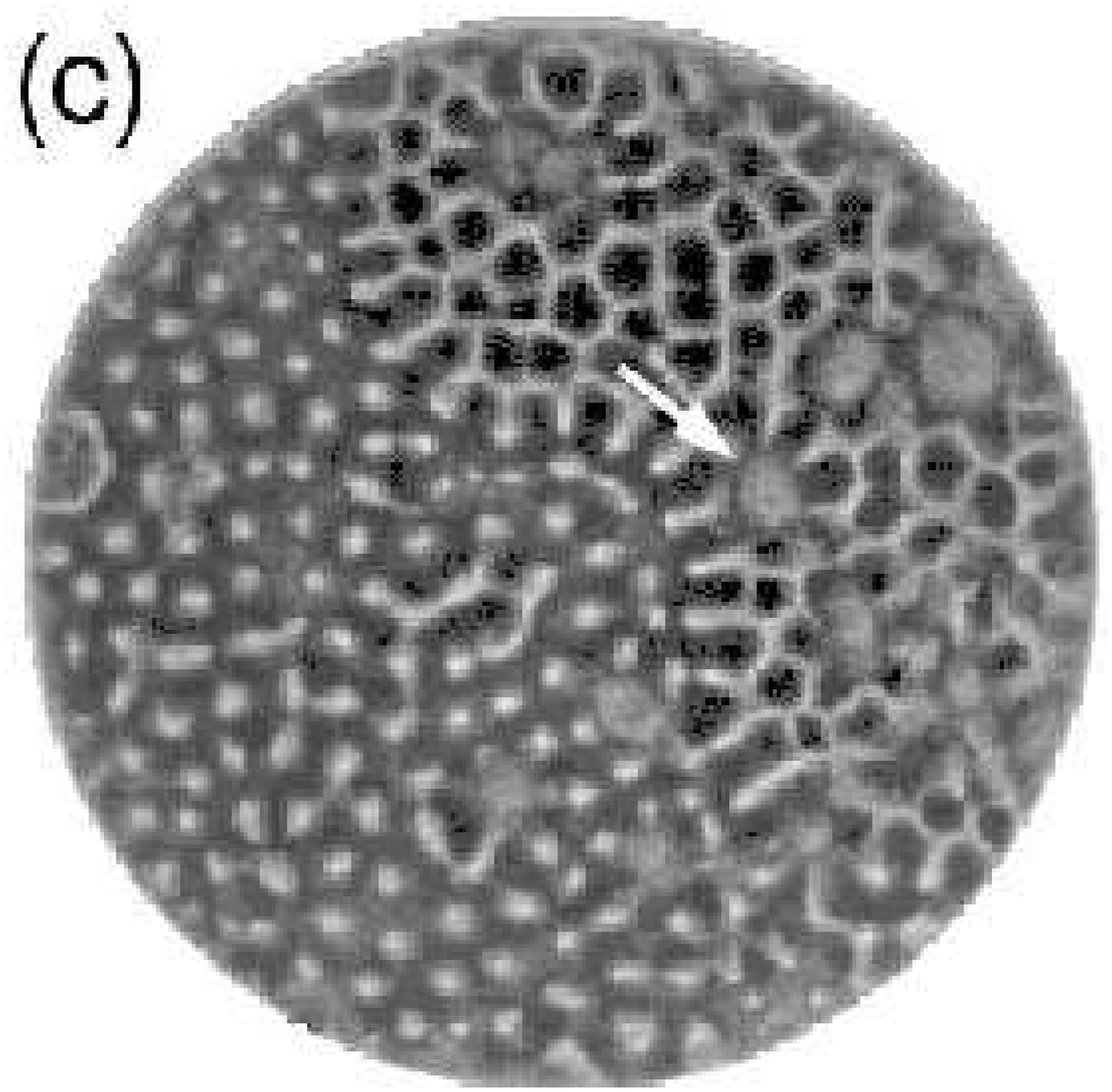}
\hskip .01\columnwidth \epsfxsize=.49\columnwidth \epsffile{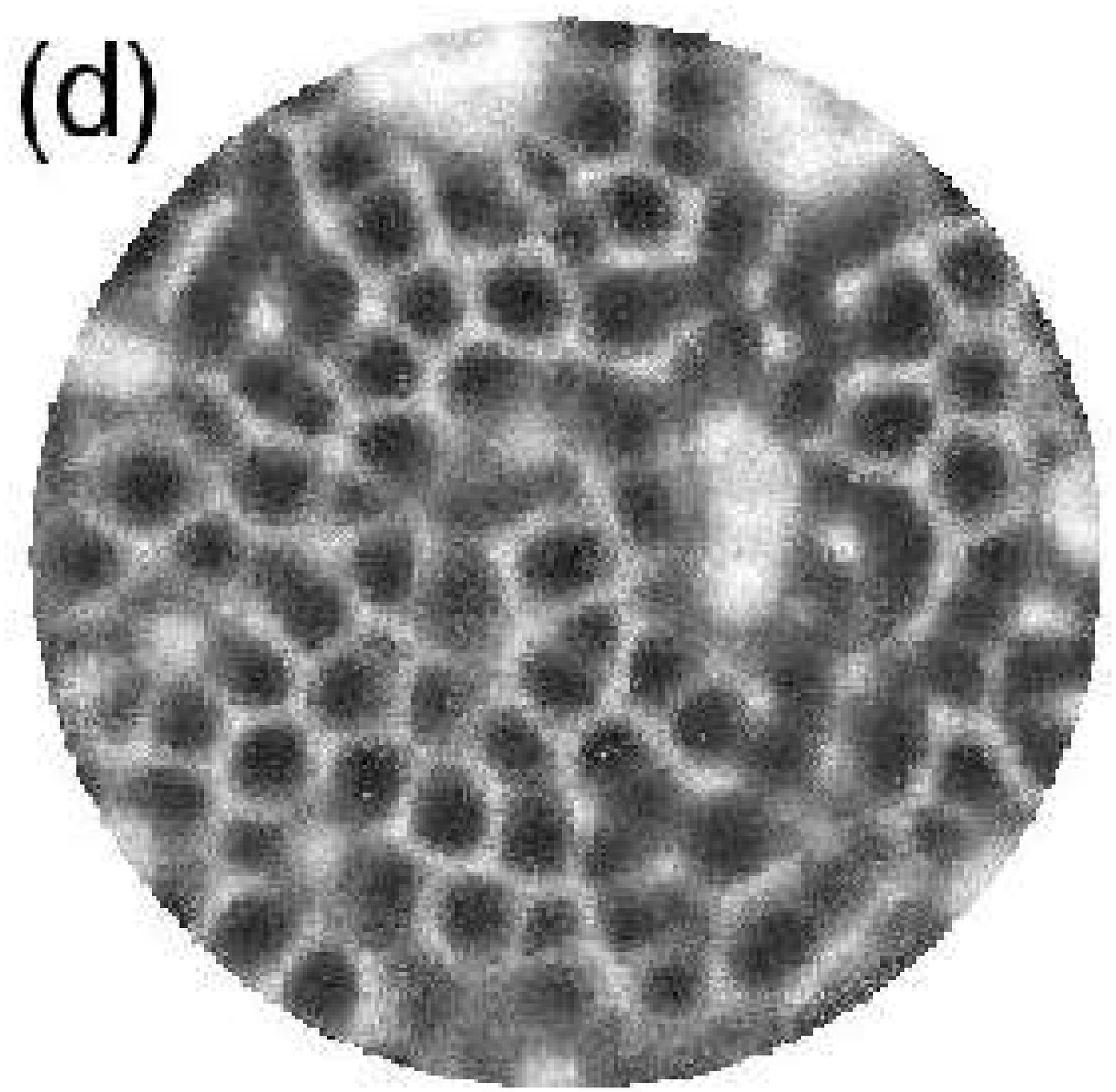}}
\centerline{{\sf Randomly moving labyrinths}}
\centerline{\epsfxsize=.49\columnwidth \epsffile{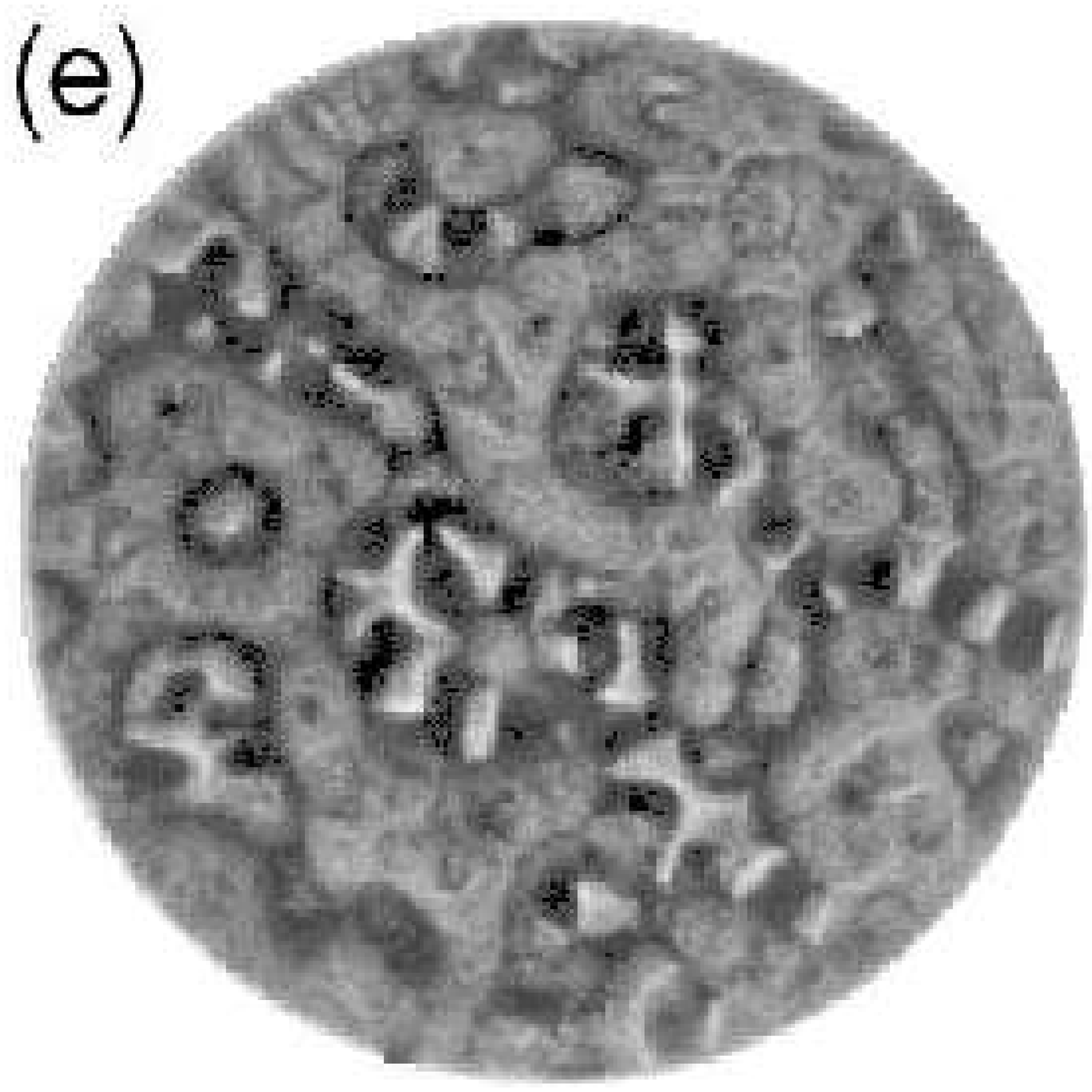}
\hskip .01\columnwidth \epsfxsize=.49\columnwidth \epsffile{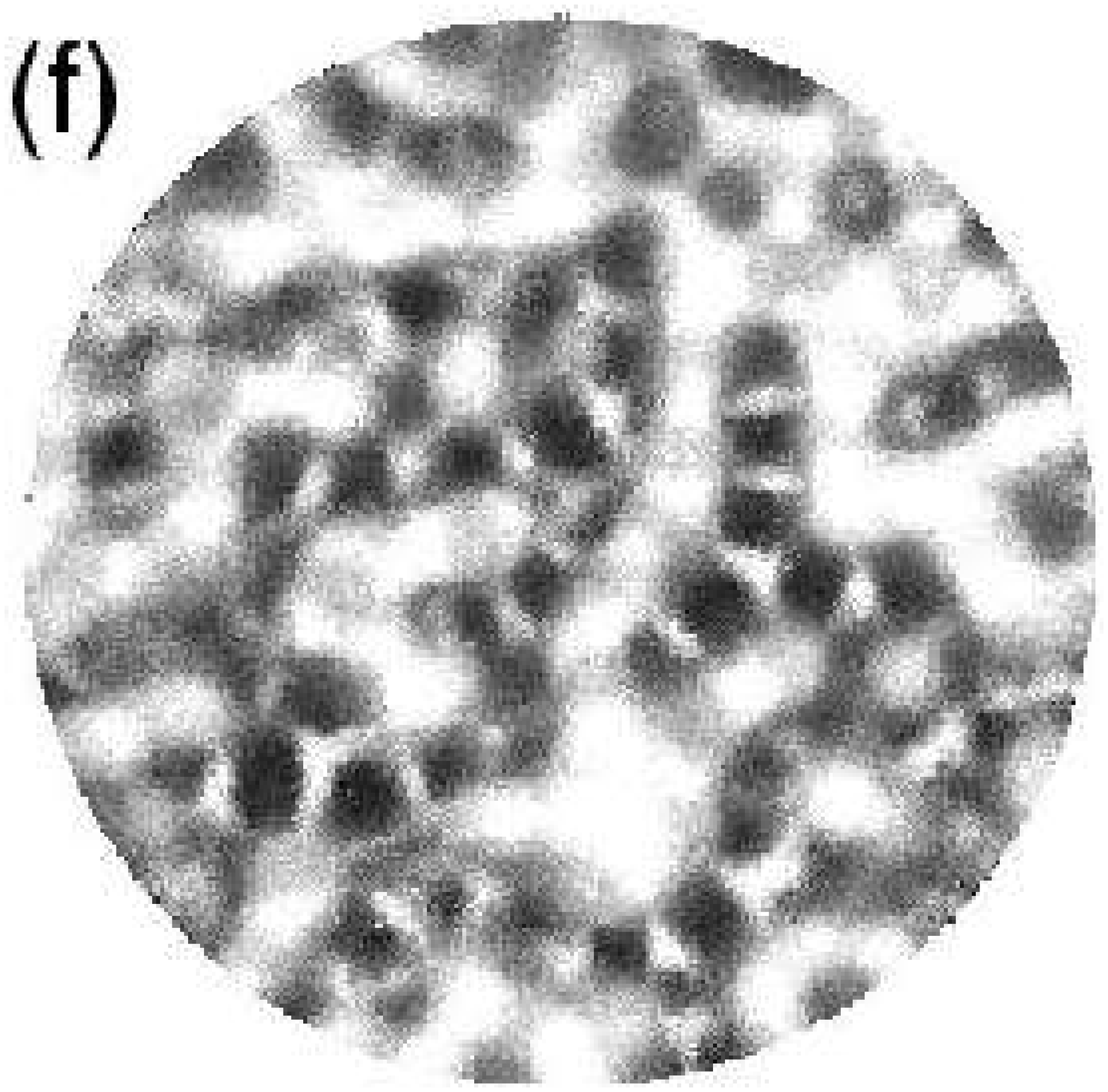}}
\caption{
Top views obtained from the experiments, and the mean height
as a function of the position obtained from 3D simulations.
In (a) and (c), there is a phase kink near the middle, 
while there is not in (b) or (d).
Kinks are easily created in the experiment due to the side wall friction
or tilt of the layer, but they are easily eliminated in the simulation;
see Sec. IV.
One phase bubble is indicated by a white arrow in (c);
in (d) the phase bubbles are the white areas,
because they reach their maximum height at this moment
and patterns superposed on them are nearly flat.
The gray scale in the experimental images indicates the intensity of
the reflected light, which is a measure of the gradient of the surface.
The gray scale in the simulation images is $\langle z\rangle(x,y)$, increasing
from black to white, where $\langle F\rangle(x,y)$ is an averaged value of $F$ 
over the particles located at $(x,y)$;
$\langle z\rangle(x,y)$ is called the mean height field.
A circular container of diameter $L = 847\sigma$ is used in the experiments,
and a circular container of $L = 294\sigma$ is simulated. 
The parameters ($\Gamma, f^*, N$) are (a) (7.1,1.0,10), (b) (7.0,0.85,8),
(c) (7.3,1.0,10), (d) (7.2,0.85,8), (e) (8.0,1.2,15), and (f) (8.9,0.85,8).
}
\end{figure}
\pagebreak

\begin{figure}
\centerline{\epsfxsize=.49\columnwidth \epsffile{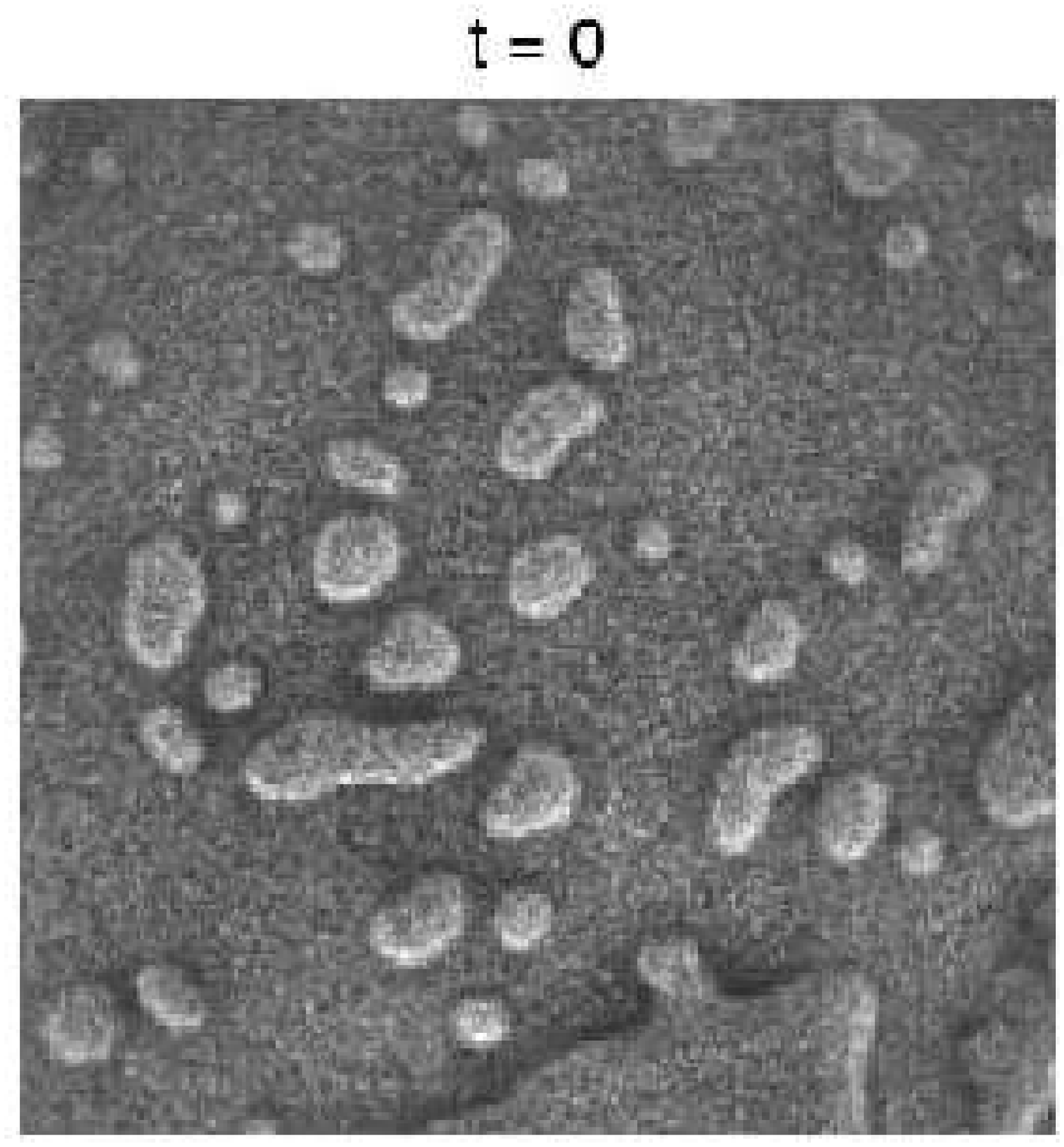} 
\hskip .01\columnwidth \epsfxsize=.49\columnwidth \epsffile{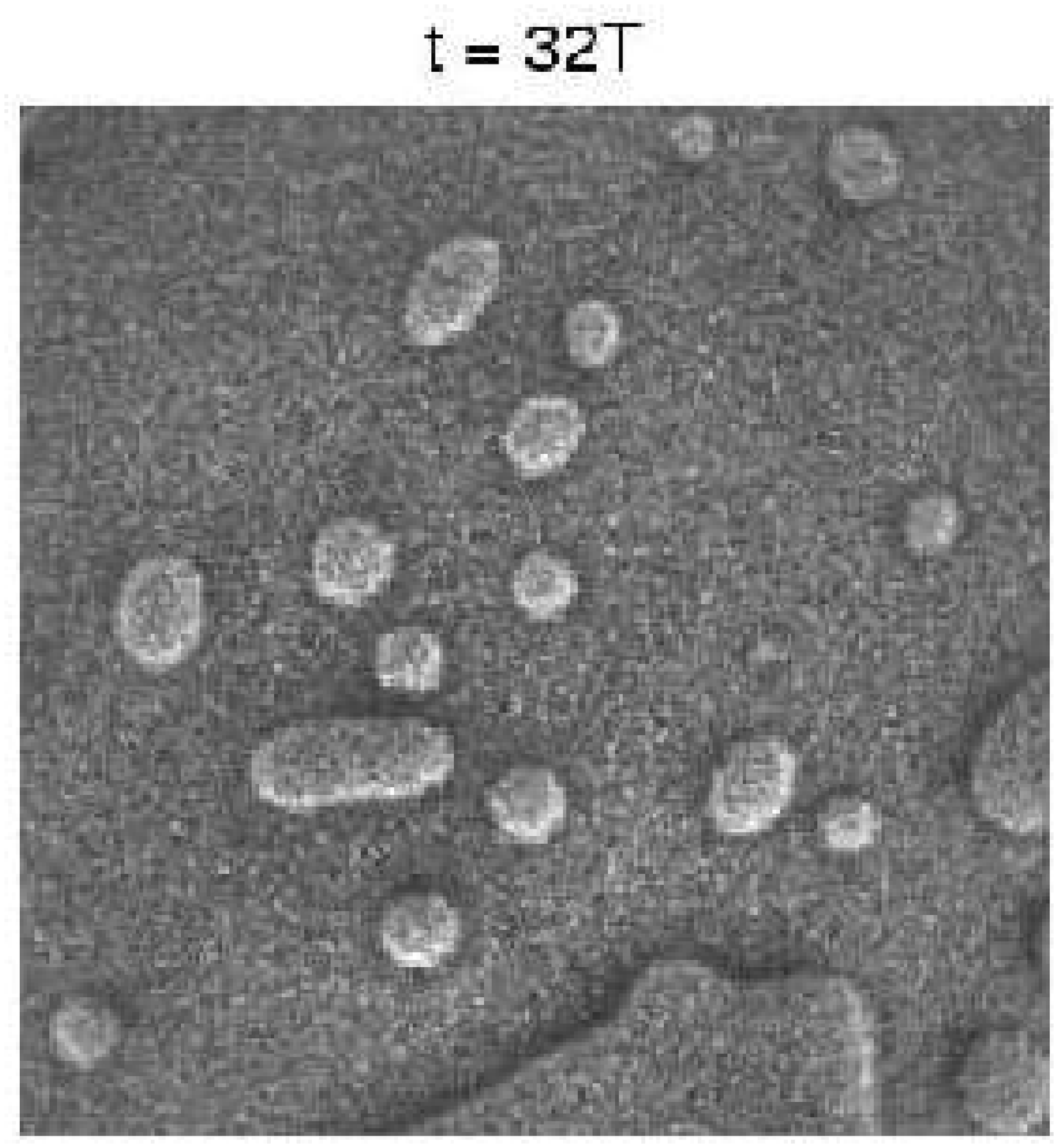}}
\centerline{\epsfxsize=.49\columnwidth \epsffile{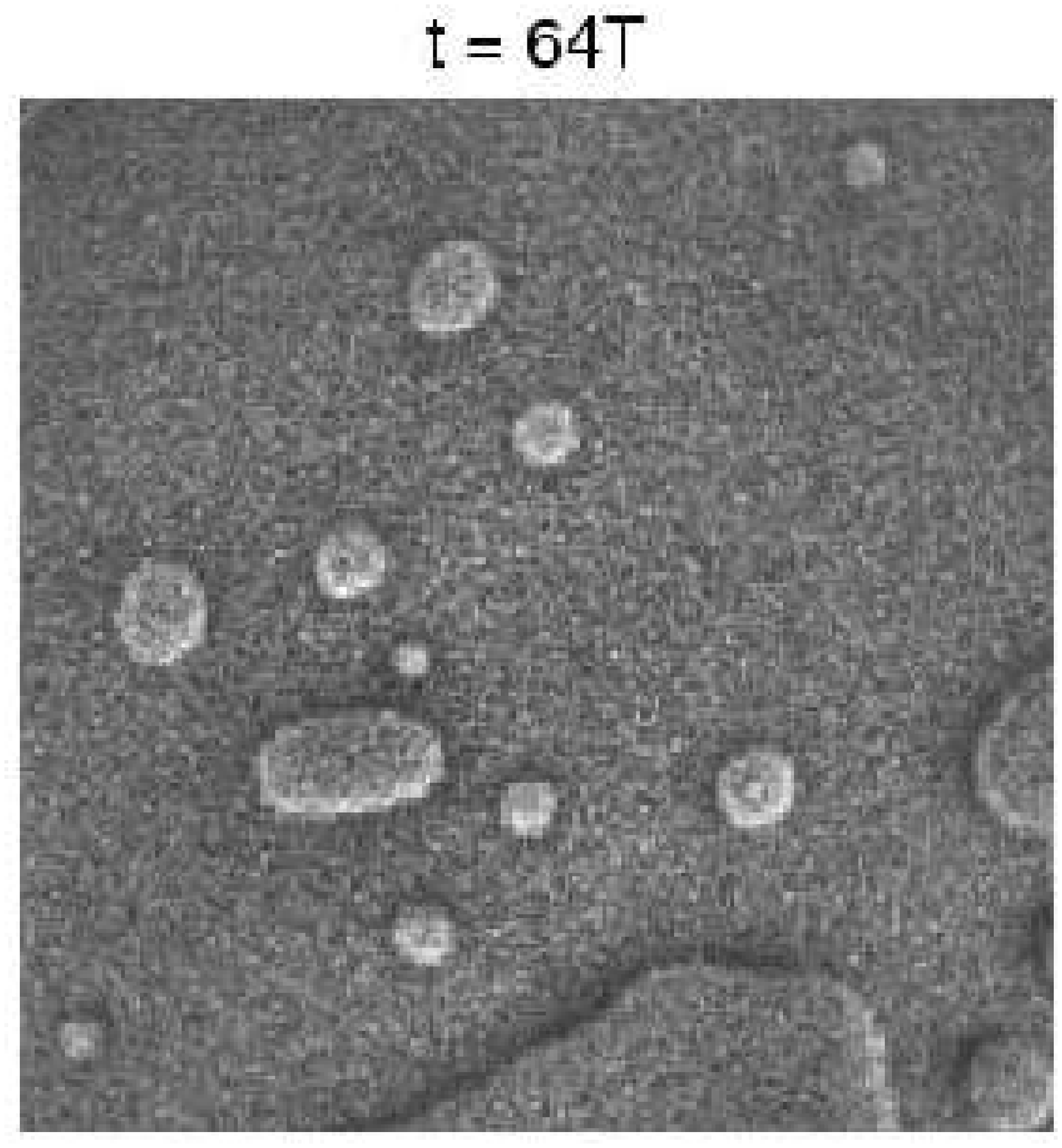} 
\hskip .01\columnwidth \epsfxsize=.49\columnwidth \epsffile{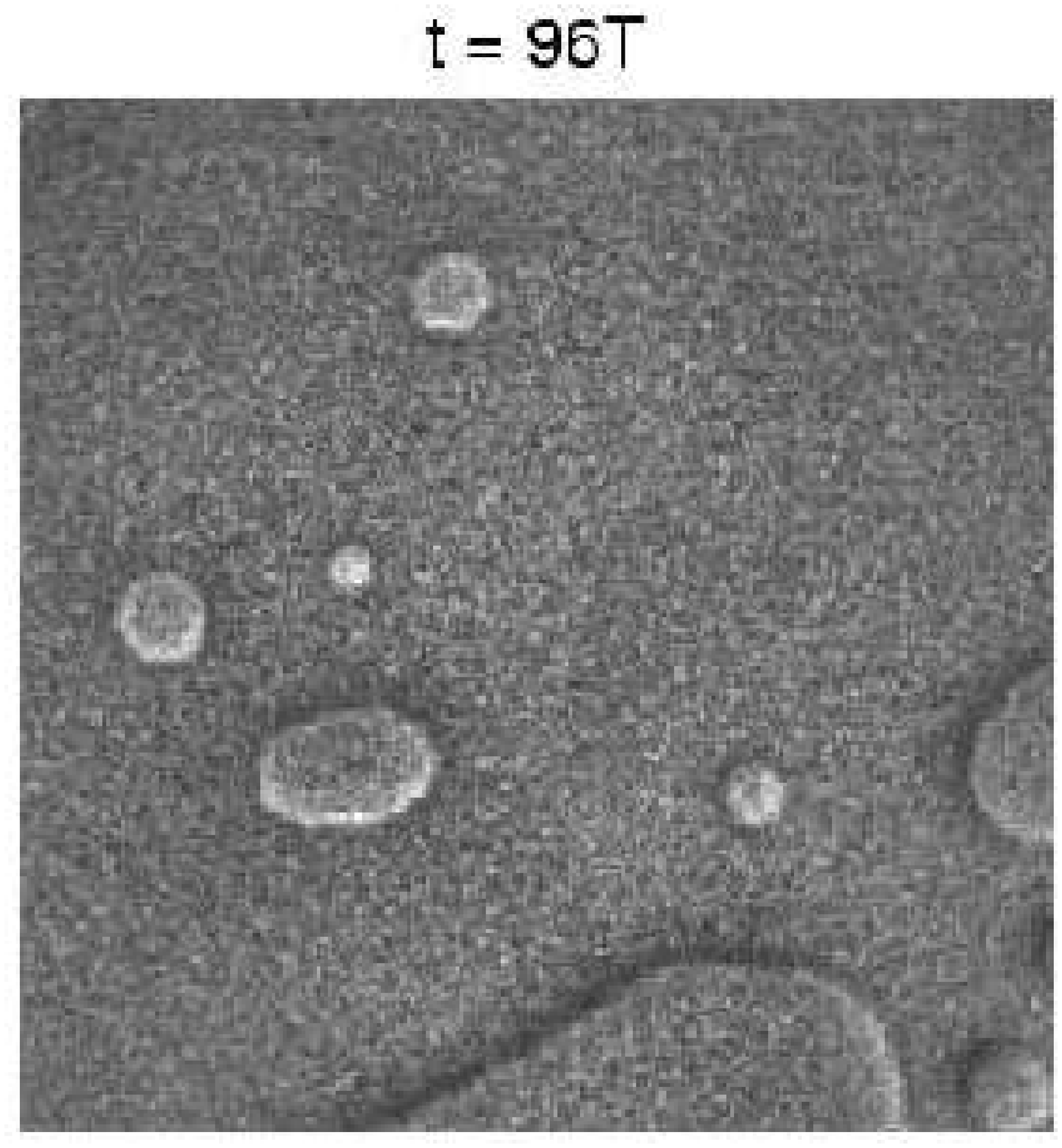}}
\caption{A sequence of phase bubbles in $f/2$ flat pattern,
obtained in the experiment.
The bubbles shrink as if they had a surface tension.
To create these phase bubbles, $\Gamma$ was suddenly decreased from $9.0$
to the $f/2$ flat pattern regime, $\Gamma = 4.5$ ($f^* = 0.6$).
Experimental setup is the same as in Fig. 1.
$400\sigma \times 400\sigma$ portion of the layer is shown.
}
\end{figure}
\pagebreak

\begin{figure}
\epsfxsize=.98\columnwidth
\centerline{\epsffile{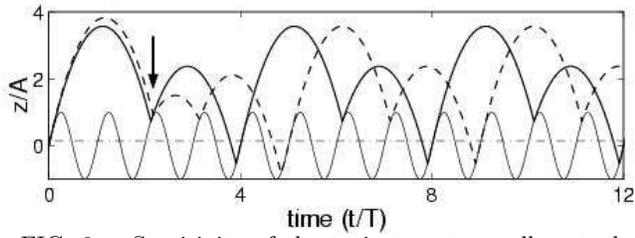}}
\caption{
Sensitivity of the trajectory to small perturbations is illustrated by
perturbed (dashed line) and unperturbed (solid line) trajectories
of a completely inelastic ball in the $f/4$ hexagonal pattern regime
($\Gamma = 7.0$).
For the perturbed trajectory, we increased the initial take-off velocity
by $3\%$ to delay the collision.
At the next collision (indicated by an arrow), the perturbed ball collided with
the container because its take-off velocity was too small, and its trajectory
became out of phase with that of unperturbed one.
}
\end{figure}
\pagebreak

\begin{figure}
\epsfxsize=.98\columnwidth
\centerline{\epsffile{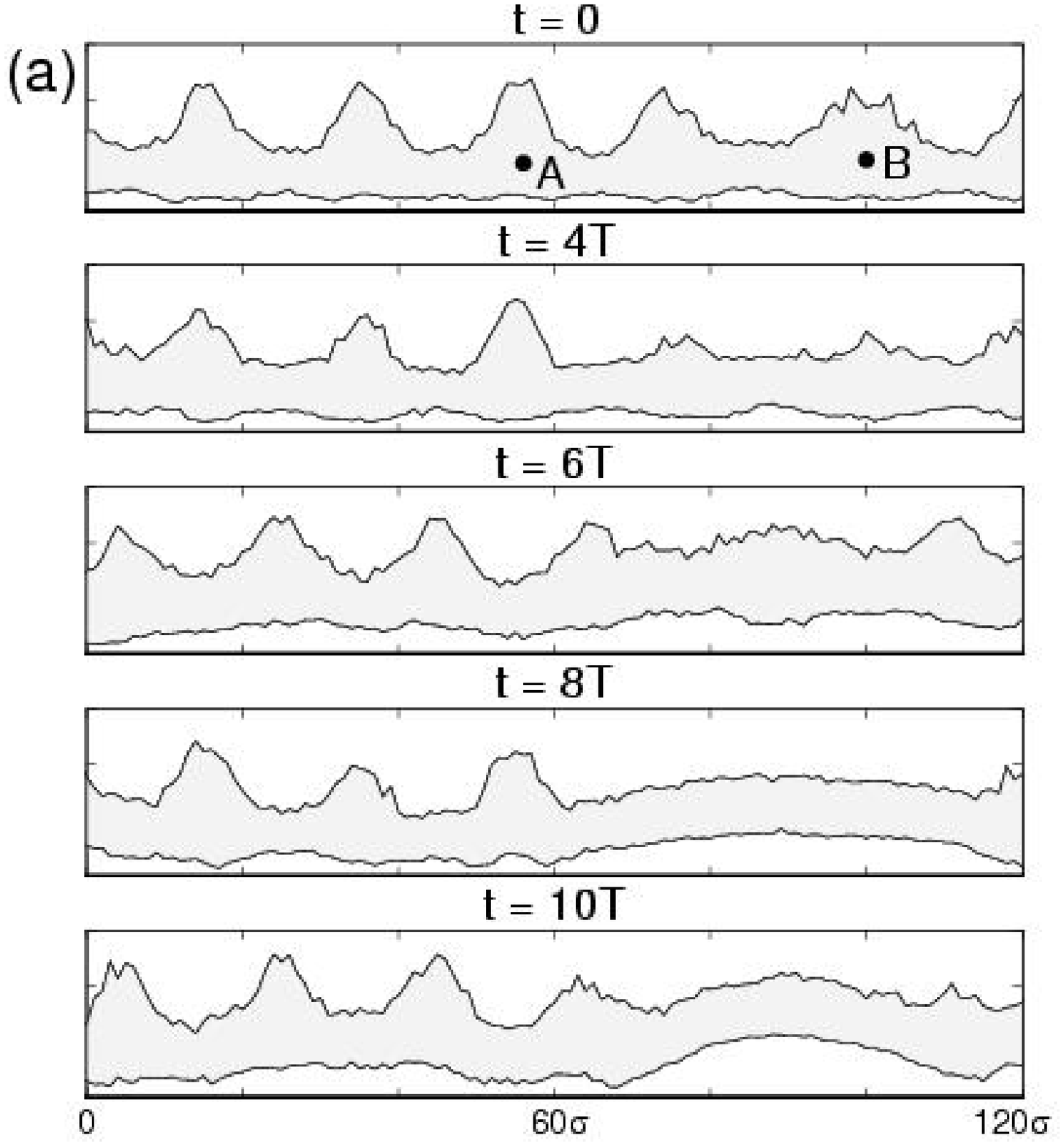}}
\epsfxsize=.98\columnwidth
\centerline{\epsffile{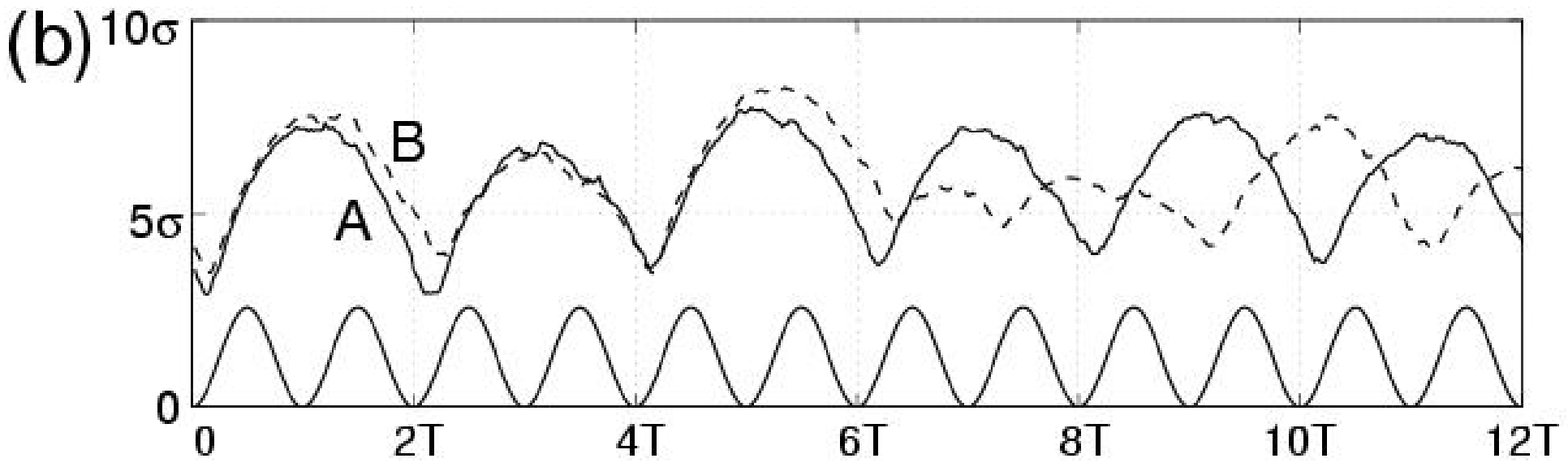}}
\caption{(a) A sequence of side views of a quasi-2D layer showing
the nucleation of a phase bubble around {\bf B}, obtained in a simulation of
a quasi-2D layer of size $200\sigma \times 10\sigma$,
for $\Gamma = 7.2$, $f^* = 0.9$, and $N = 6$. 
(b) The trajectories of the center of mass at two horizontal locations,
{\bf A} and {\bf B} in (a), as a function of time.
The sinusoidal function shown in (b) is the trajectory of the plate
of the container.
At $t = 0$ in (a), the bottom of the layer has a short length scale
deformation slaved to the pattern on the surface of the layer.
This undulation at the bottom grows, as shown in the next two successive
frames in (a) ($t = 4T$ and $6T$ in (b)).
At $t = 6T$, the collision of {\bf B} is delayed due to the undulation;
at this collision, a portion of the layer around {\bf B} collides with
the container when the plate nearly reaches its maximum height,
and the take-off velocity becomes too small to fly
the next two cycles over the container.
As a result, {\bf B} collides with the container at the very next cycle,
at $t \sim 7T$, and it becomes nearly $\pi$ out of phase with the rest
of the layer.
This mechanism is the same as that in Fig. 6.
}
\end{figure}
\pagebreak

\begin{figure}
\centerline{\epsfxsize=.98\columnwidth \epsffile{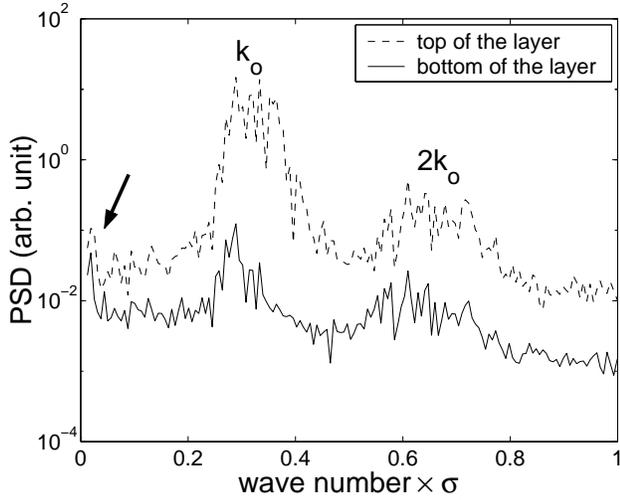}}
\caption{
Power spectral densities (PSD's) of the top and the bottom of a quasi-2D
layer for $\Gamma = 6.6, f^* = 0.9$, and $N = 6$, obtained in simulations
of a quasi-2D layer of size $1000\sigma \times 10\sigma$ having more than
50 wavelengths; there is an additional peak (indicated by an arrow) at a small
wavenumber as well as the primary peaks at the wavenumber of the pattern
(at $k_o = 2\pi\sigma /\lambda$, where $\lambda$ is the wavelength of the pattern)
and its subharmonic. $\Gamma_{pb} = 6.7$ for this simulation.
PSD's were measured when the amplitude of the pattern was fully developed,
just before the layer collided with the container.
}
\end{figure}
\pagebreak

\begin{figure}
\centerline{\epsfxsize=.49\columnwidth \epsffile{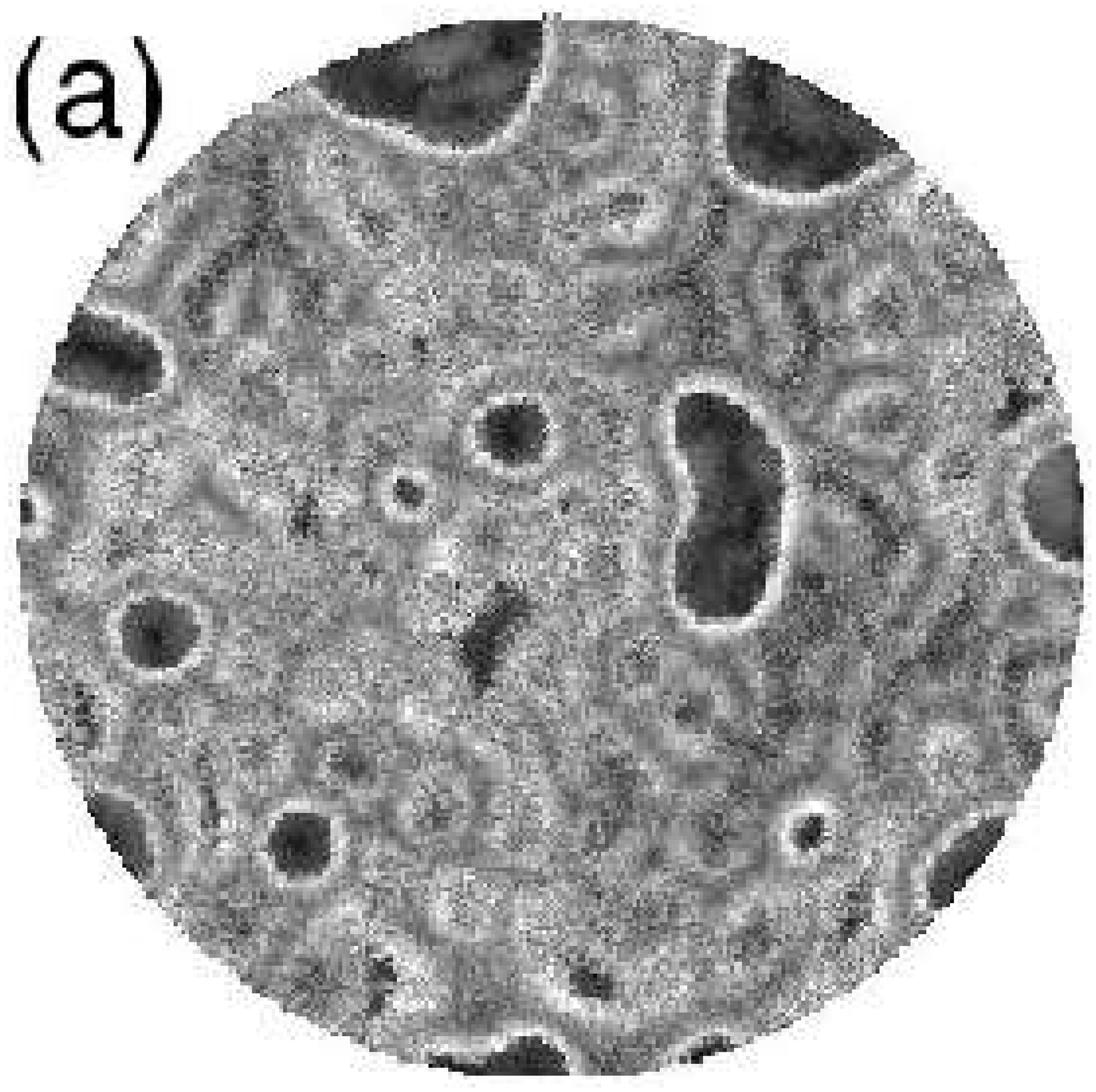} 
\hskip .01\columnwidth \epsfxsize=.49\columnwidth \epsffile{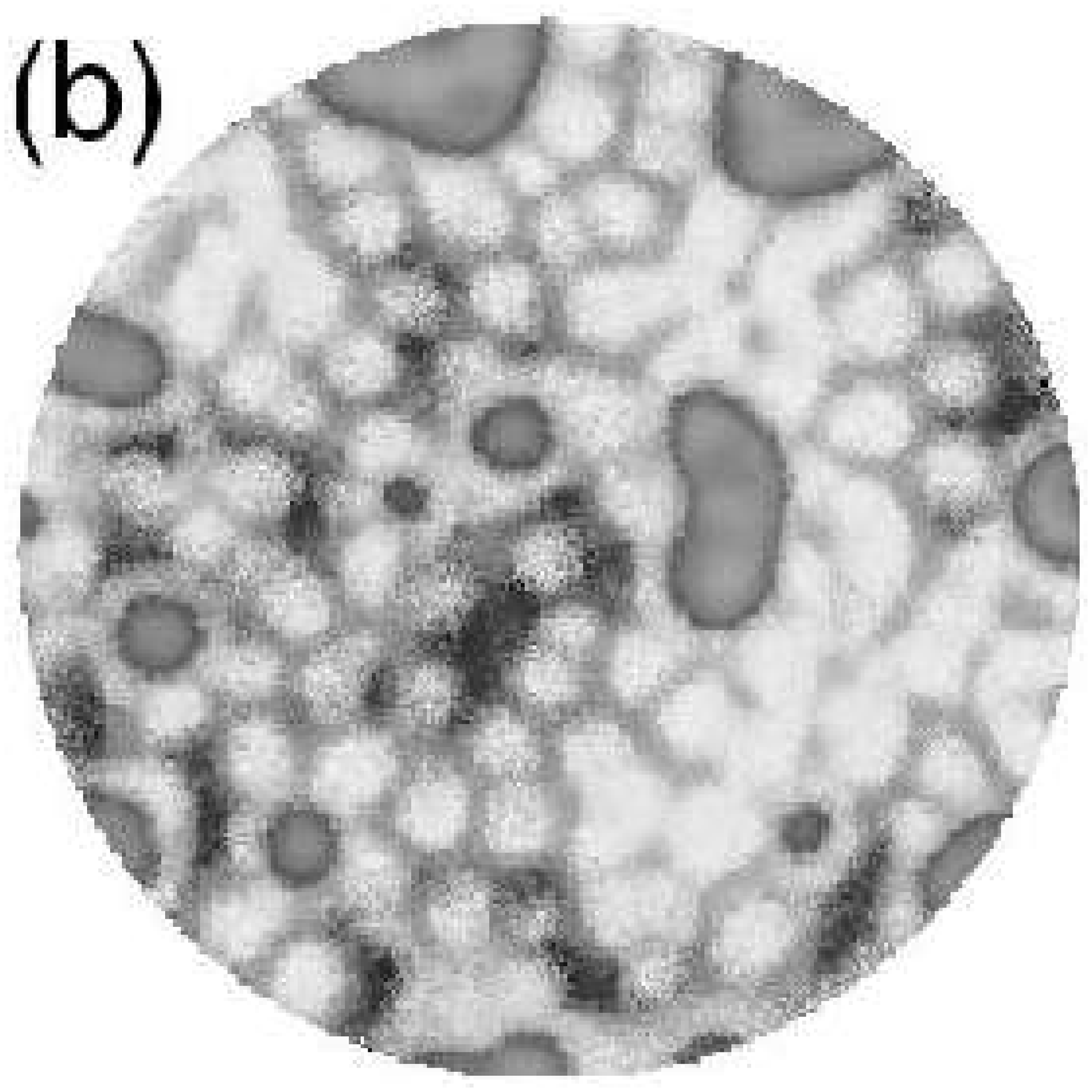}}
\caption{
(a) Horizontal momentum field $\langle \sqrt{p_x^2+p_y^2}\rangle(x,y)$,
which has its maximum values (indicated white) along the boundary
of phase bubbles; there is a significant momentum flux perpendicular
to the boundary (the kink).
(b) Vertical momentum field $\langle p_z\rangle(x,y)$, black being
the maximum downward, and white being the maximum upward.
Fluctuation in vertical velocities (blackish areas) is due to
the large scale undulation of the layer, discussed in Sec. IV.
These are obtained from the same simulation data as in Fig. 4(d).
}
\end{figure}
\pagebreak

\begin{figure}
\epsfxsize=.98\columnwidth
\centerline{\epsffile{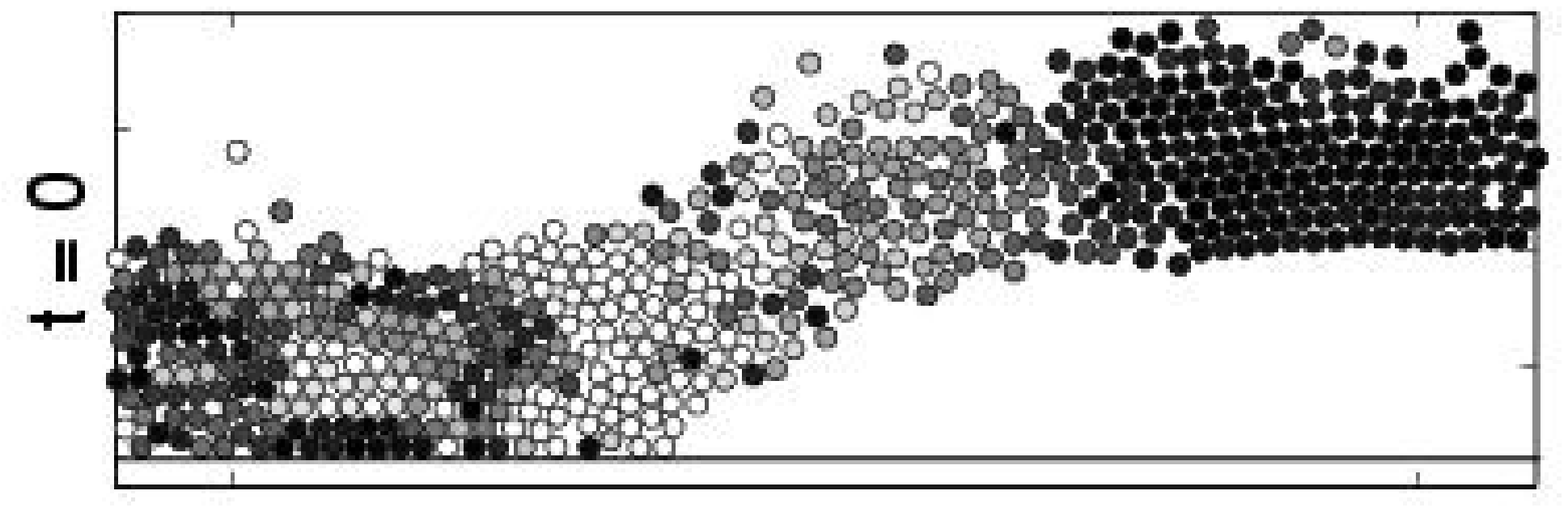}}
\epsfxsize=.98\columnwidth
\centerline{\epsffile{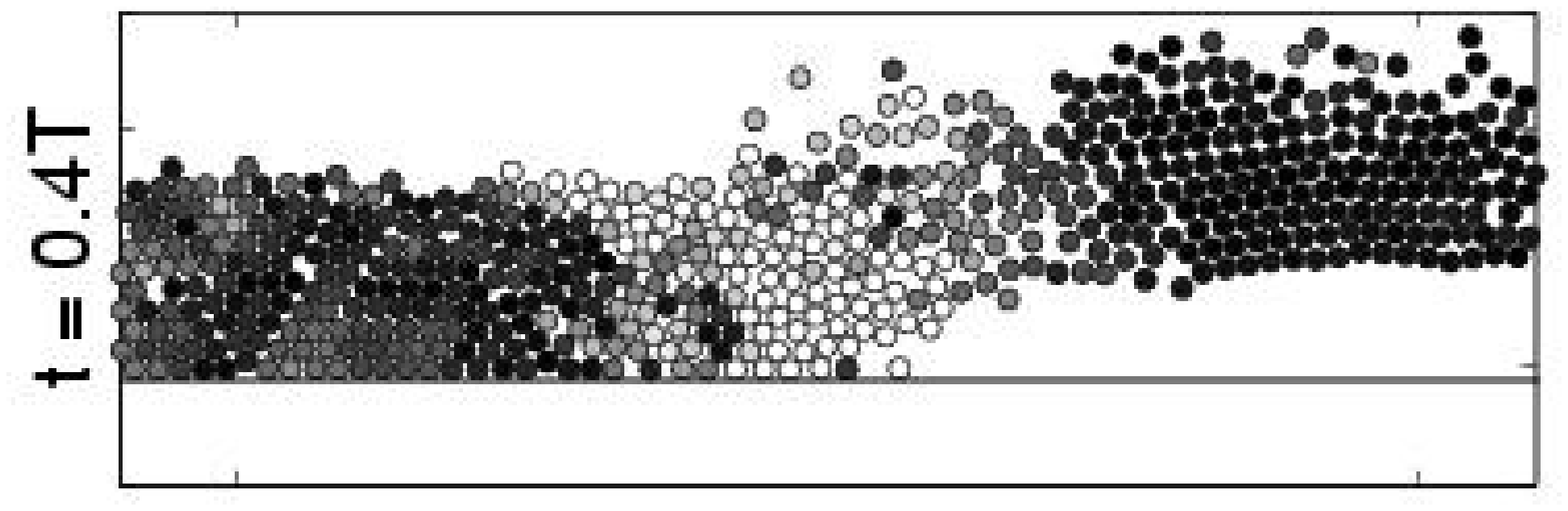}}
\epsfxsize=.98\columnwidth
\centerline{\epsffile{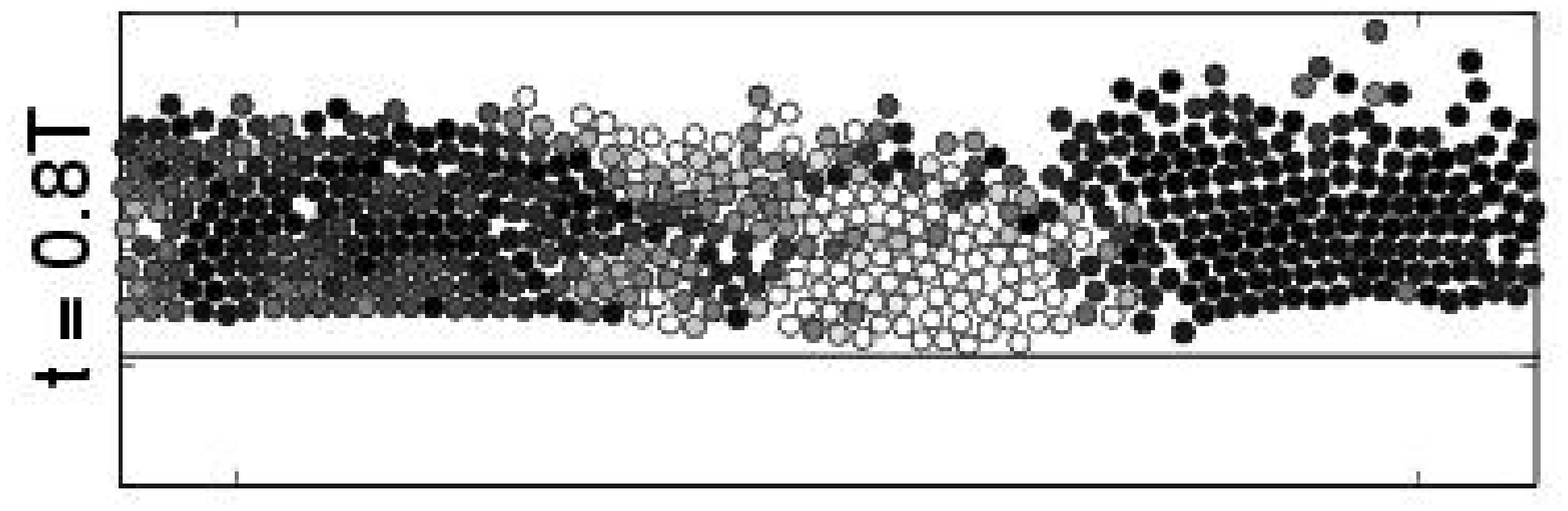}}
\epsfxsize=.98\columnwidth
\centerline{\epsffile{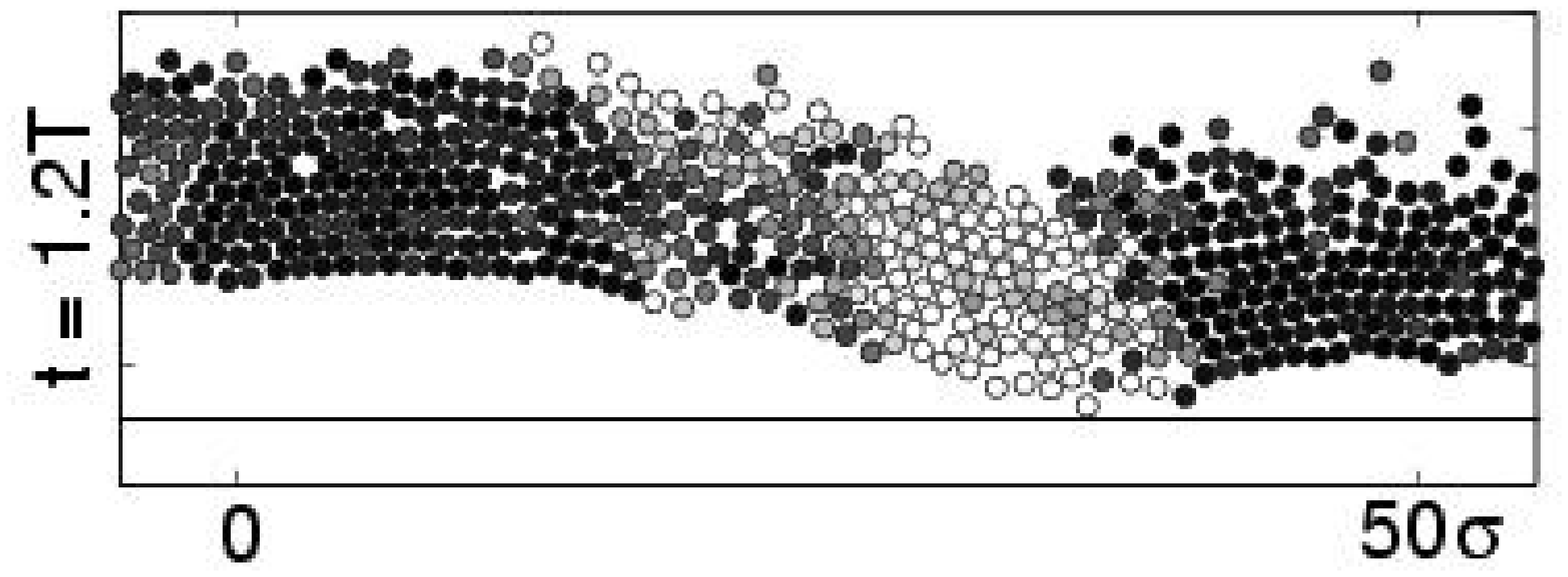}}
\caption{A sequence of an $f/2$ flat pattern with a kink,
obtained from the simulation of a 2D layer for
$\Gamma = 5.2, f^* = 0.6$, and $N = 9$.
Each circle corresponds to a particle.
The gray scale of the particles indicates the magnitude of
the horizontal momentum $|p_x|$, increasing from black to white.
As the layer is pushed up by the plate, a density wave forms,
which initiates a momentum transfer propagating rightward
(see text for details).
}
\end{figure}
\pagebreak

\begin{figure}
\centerline{\epsfxsize=.49\columnwidth \epsffile{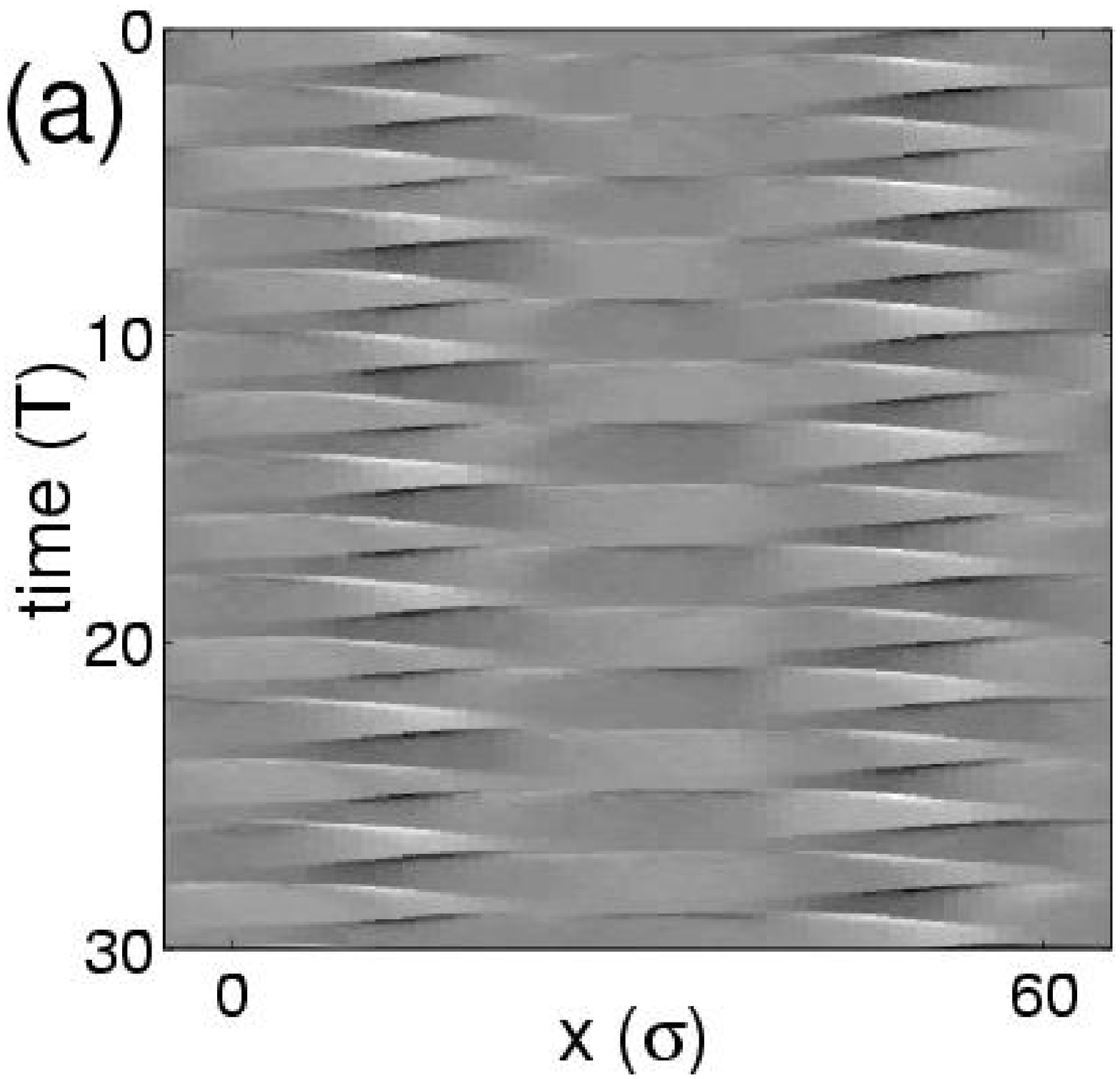} 
\hskip .01\columnwidth \epsfxsize=.49\columnwidth \epsffile{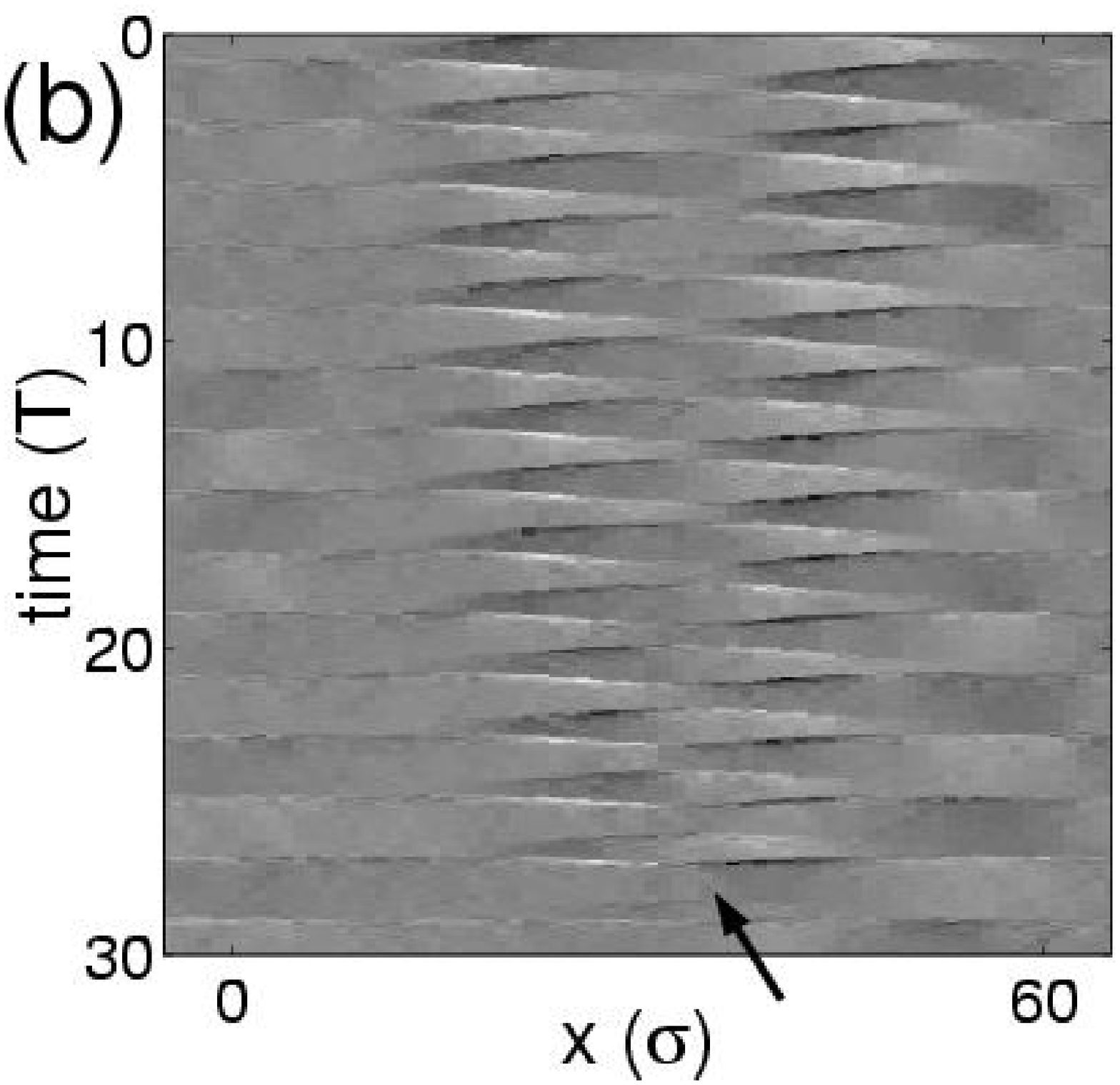}}
\caption{Space-time plots of horizontal momentum $\langle p_x\rangle(x,t)$
of a phase bubble (a) in a quasi-2D $f/2$ flat pattern,
and (b) in a 3D $f/2$ flat pattern (cross section along the diameter
of a phase bubble).
The values increase from black to white, black being the maximum leftward
and white being the maximum rightward.
A phase bubble in a quasi-2D layer oscillates symmetrically, while
in a 3D layer oscillates asymmetrically, shrink and disappear
(indicated by an arrow).
A quasi-2D layer of size $100\sigma \times 10\sigma$ is used for (a), and
the phase bubble in (b) was created by decreasing $\Gamma$ from $8.5$
to $5.2$. $f^* = 0.6$ and $N = 6$ for both.
}
\end{figure}
\pagebreak

\begin{figure}
\centerline{\epsfxsize=.49\columnwidth \epsffile{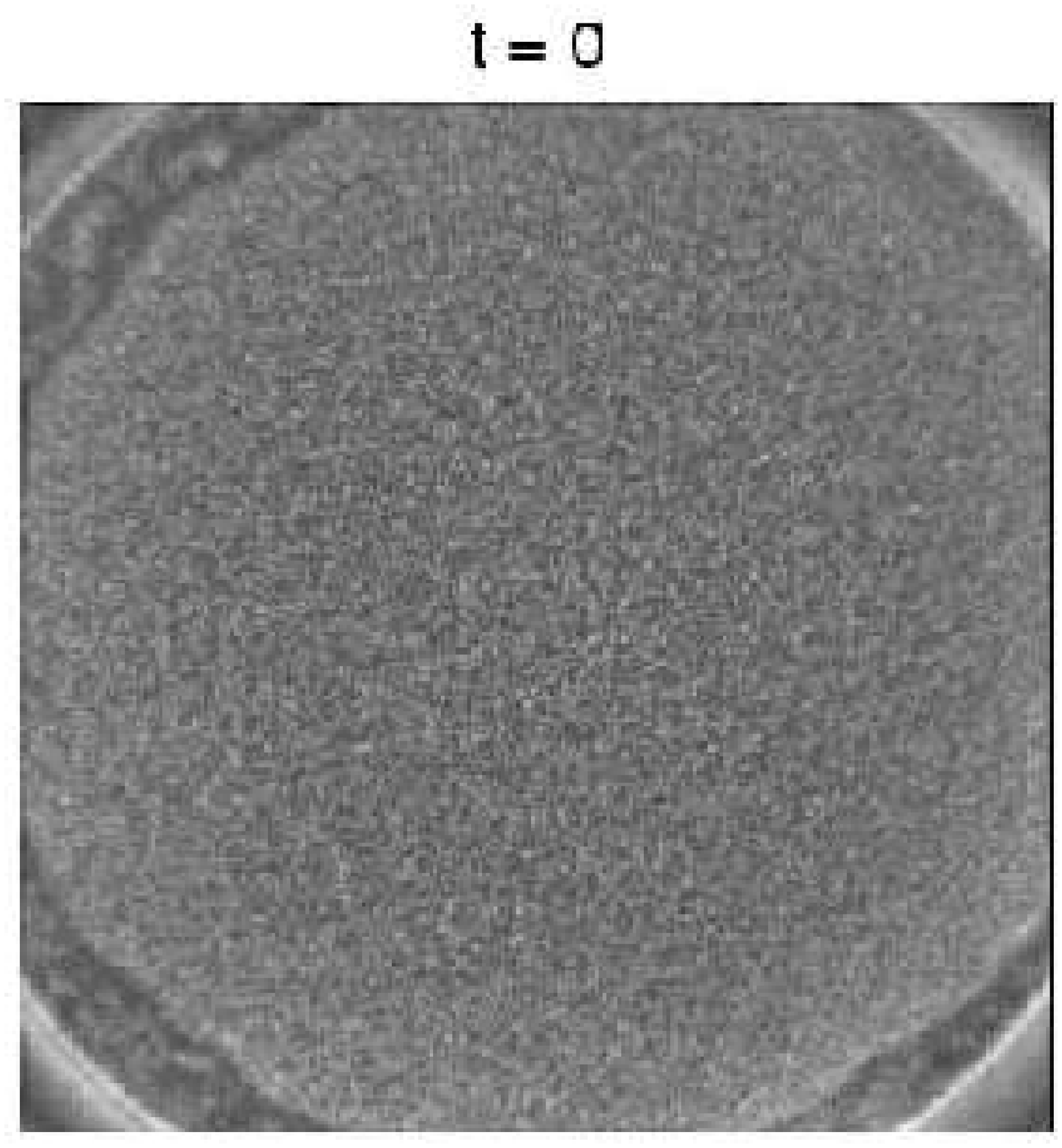} 
\hskip .01\columnwidth \epsfxsize=.49\columnwidth \epsffile{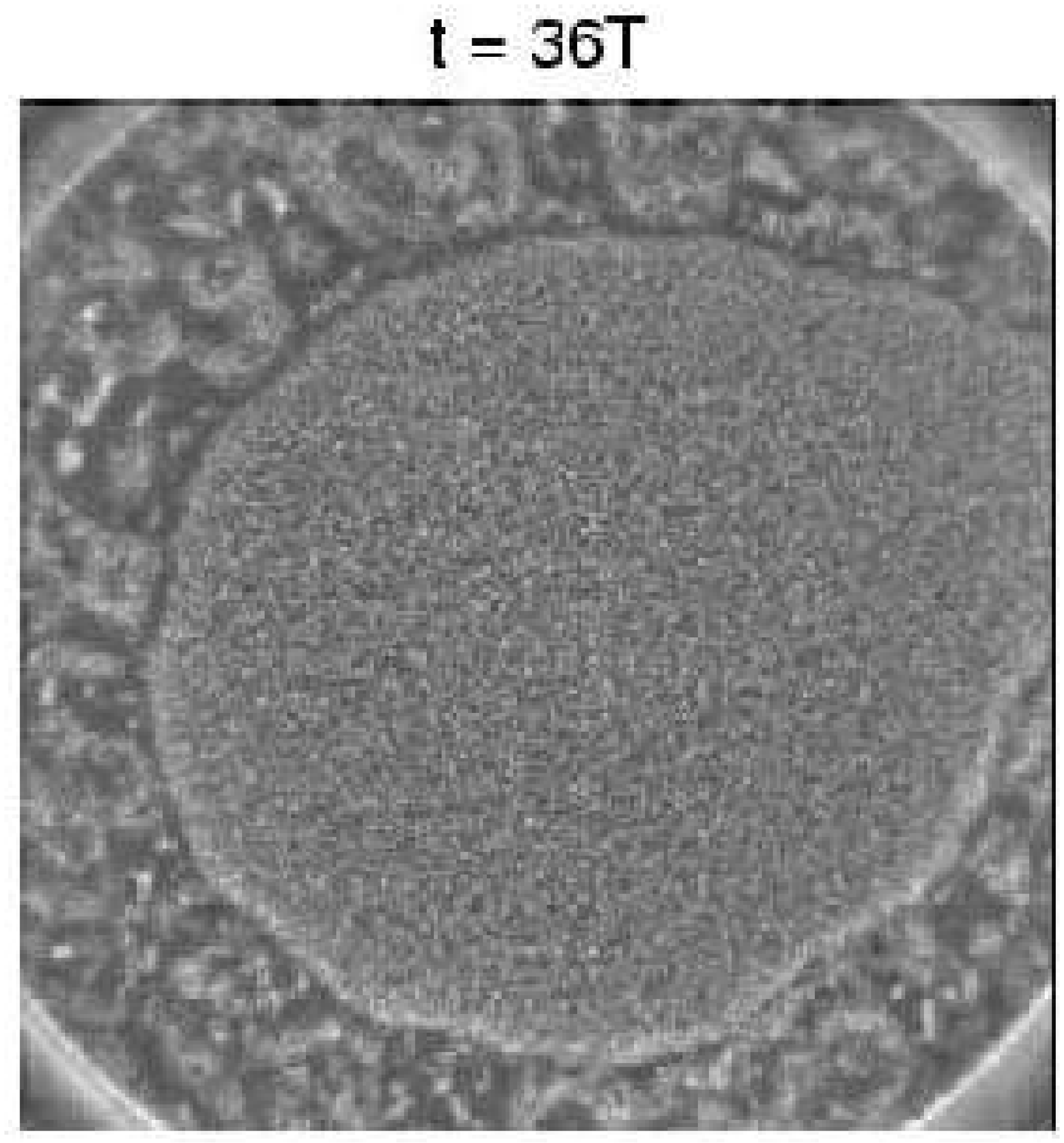}}
\centerline{\epsfxsize=.49\columnwidth \epsffile{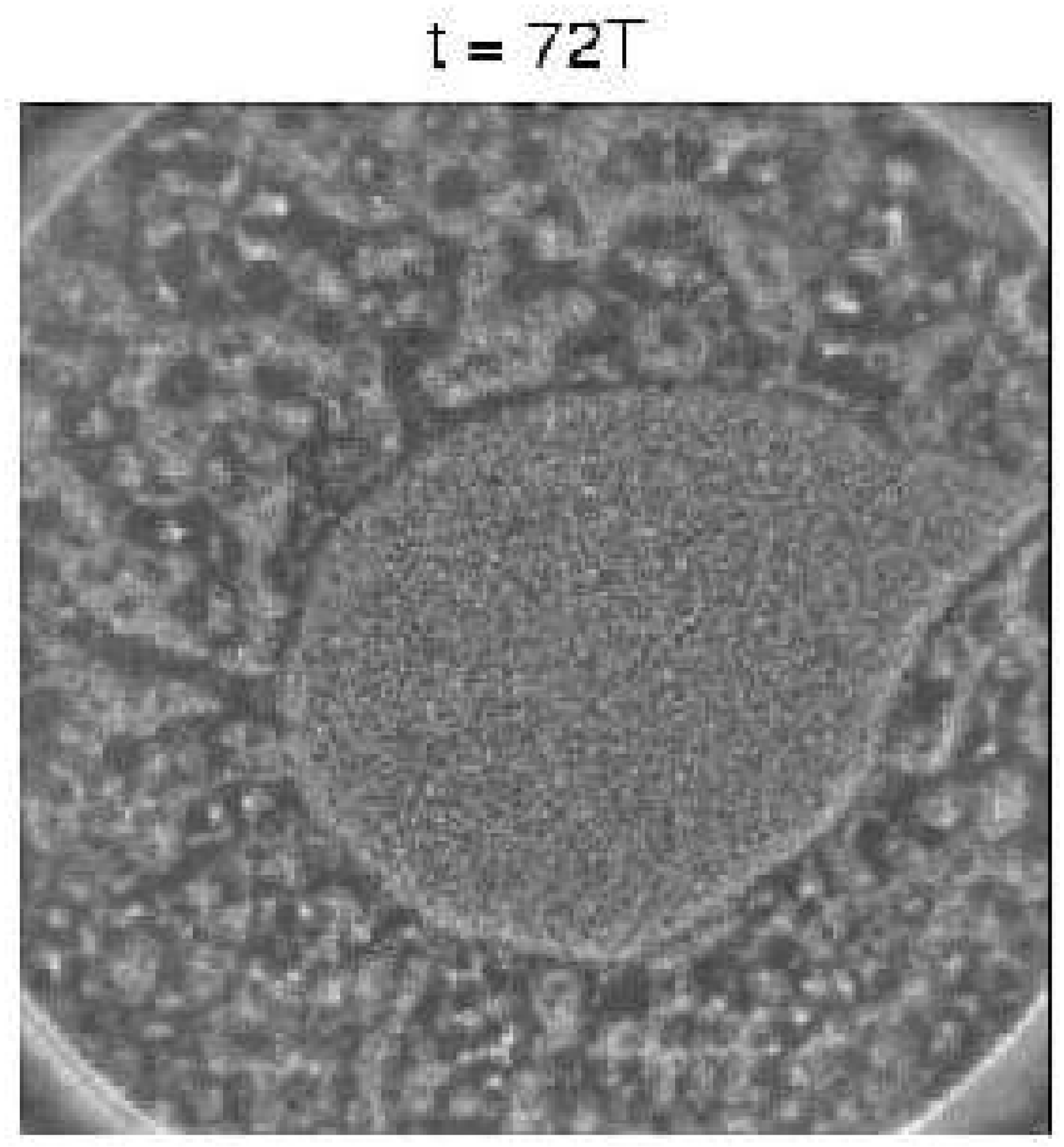} 
\hskip .01\columnwidth \epsfxsize=.49\columnwidth \epsffile{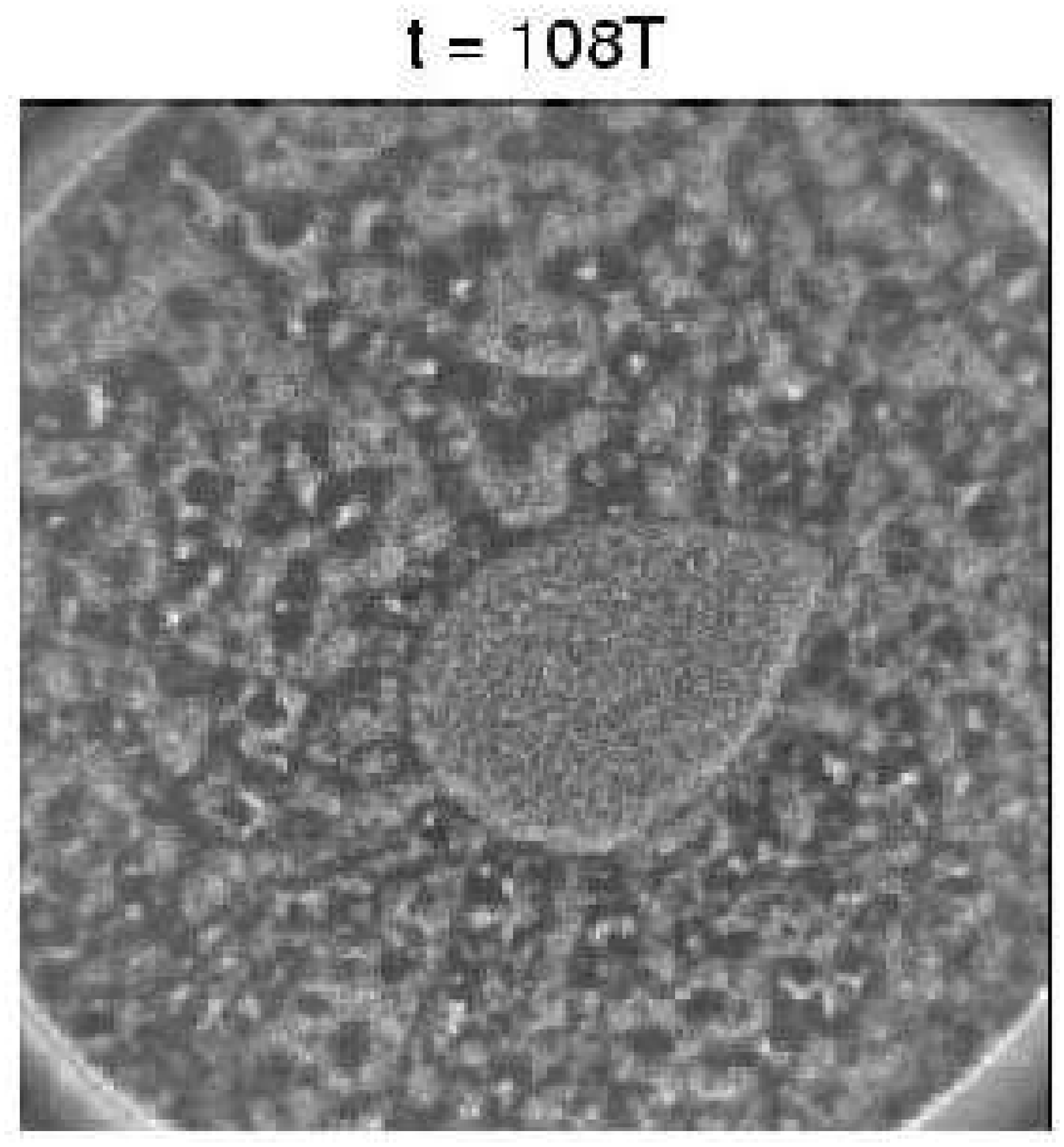}}
\caption{A sequence of a transient $f/3$ flat pattern 
obtained from the experiment.
The friction due to the side wall creates a kink, which propagates
to the center of the container in the radial direction (the same mechanism
of the shrinking of a phase bubble) and destroys the $f/3$ flat pattern.
$\Gamma$ was suddenly increased from below the onset, about $2.0$, 
to 7.8; $f^* = 0.94$ and $N = 6$.
The experiment was done with the same particles as in Fig. 1,
and $L = 847\sigma$.
}
\end{figure}
\pagebreak

\begin{figure}
\centerline{\epsfxsize=.49\columnwidth \epsffile{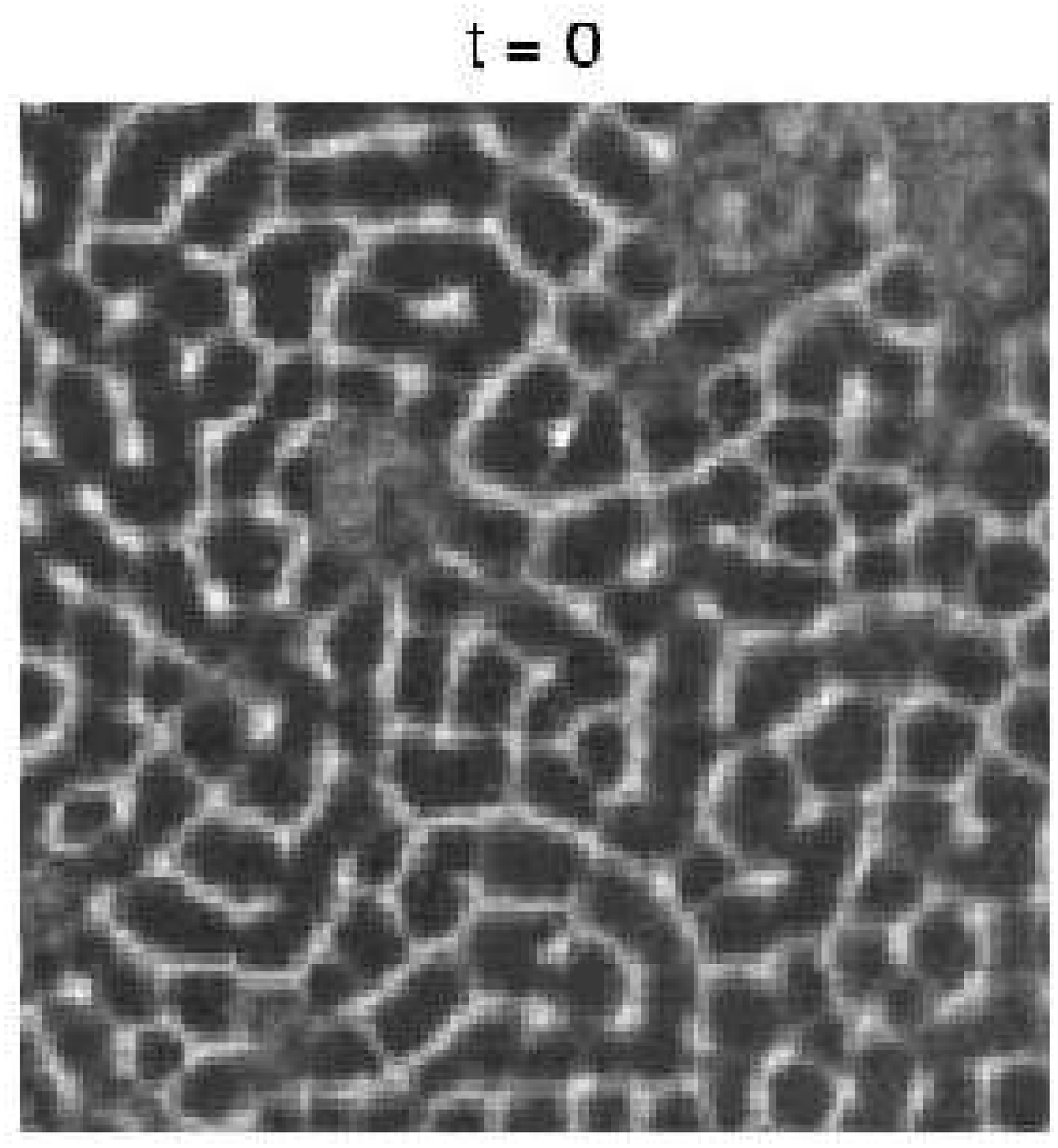} 
\hskip .01\columnwidth \epsfxsize=.49\columnwidth \epsffile{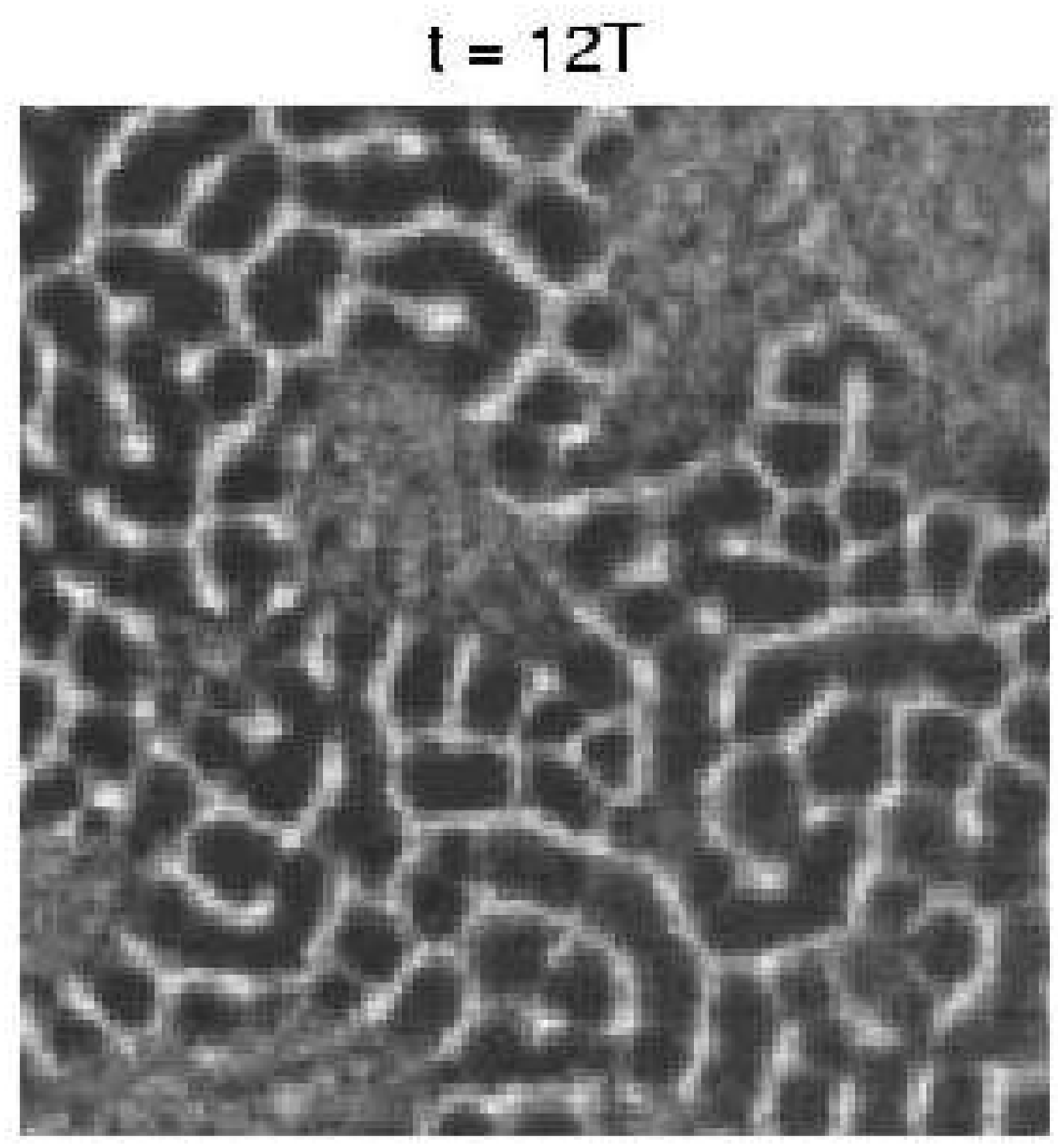}}
\centerline{\epsfxsize=.49\columnwidth \epsffile{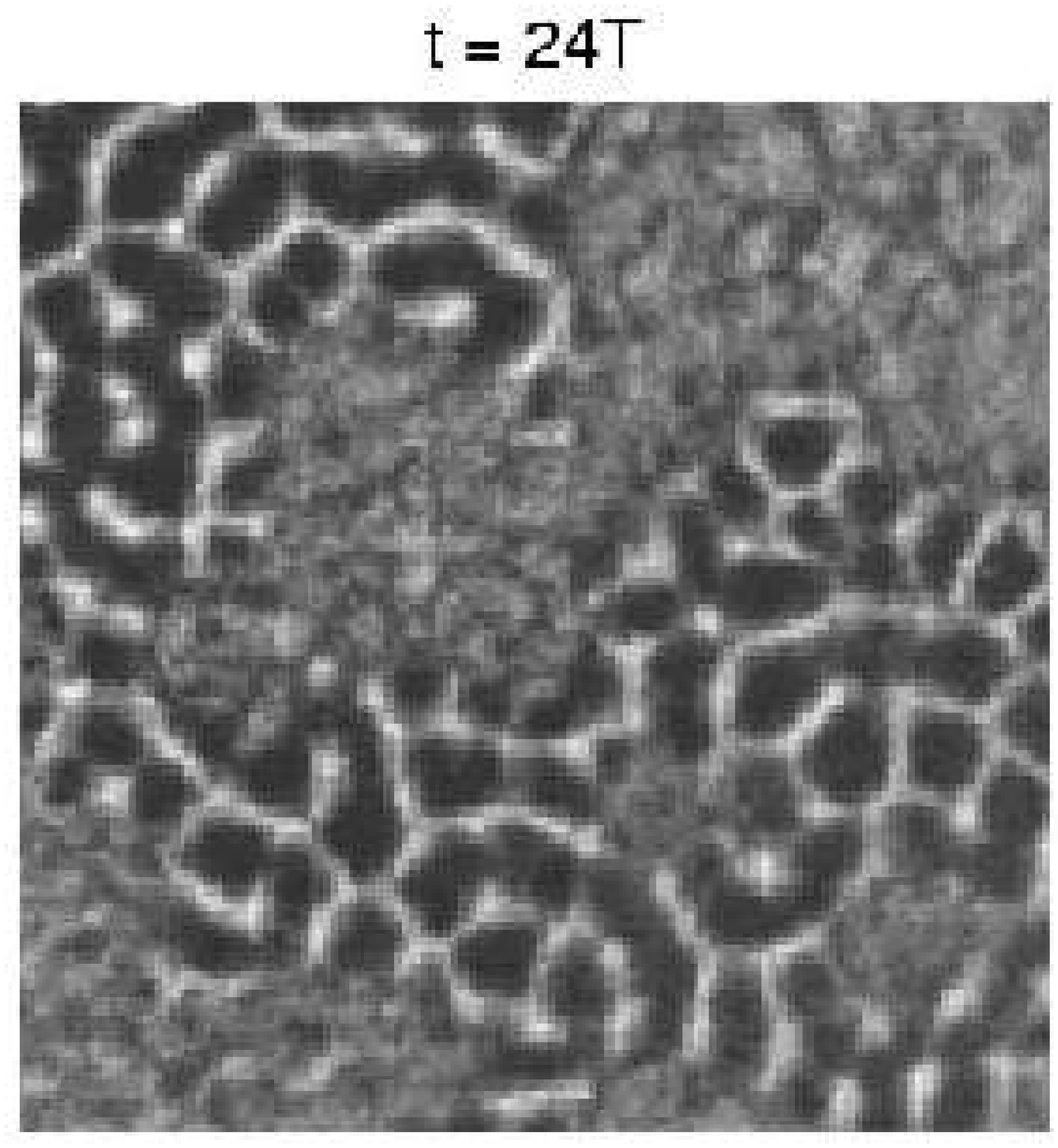} 
\hskip .01\columnwidth \epsfxsize=.49\columnwidth \epsffile{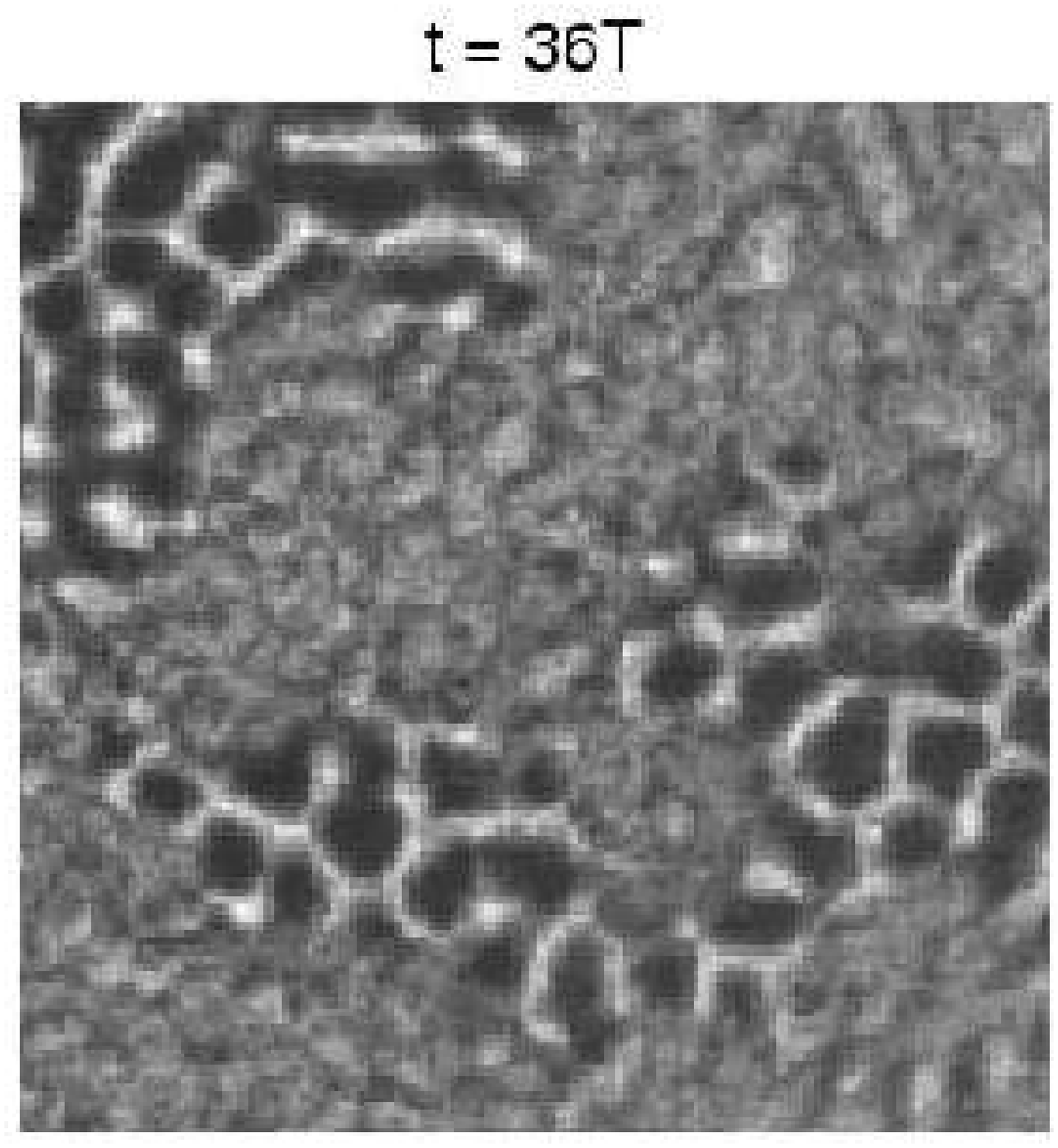}}
\caption{A sequence of a transient $f/6$ pattern,
obtained from the experiment.
The undulation of the layer creates kinks in the middle of the layer,
which destroys the $f/6$ pattern.
$\Gamma$ was suddenly increased from 2.0 (below pattern onset),
to  10 ($f^* = 1.2$ and $N = 6$).
The experiment was done with the same particles, $N$, and $L$ as in Fig. 13,
and a part of the layer of size $250\sigma \times 250\sigma$ is shown.
}
\end{figure}
\pagebreak

\end{multicols}


\begin{references}

\bibitem[*]{email}
Electronic mail: moon@chaos.ph.utexas.edu
\bibitem[\ddagger]{hlsemail}
Electronic mail: swinney@chaos.ph.utexas.edu
\bibitem[\dagger]{newaddress}
Current address: Department of Physics, City College of CUNY,
New York, NY 10031-9198
\vspace{.5cm}
\bibitem{cross93}
M. C. Cross and P. C. Hohenberg, Rev. Mod. Phys. {\bf 65}, 851 (1993).
\bibitem{amplitudefield}
V. Steinberg, E. Moses, and J. Fineberg, Nucl. Phys. B (Proc. Suppl.) 
{\bf 2}, 109 (1987);
P. Kolodner, J. A. Glazier, and H. Williams, Phys. Rev.  Lett. 
{\bf 65}, 1579 (1990).
\bibitem{rbc}
A. Pocheau, V. Croquette, and P. Le Gal, Phys. Rev. Lett. {\bf 55},
1094 (1985). 
\bibitem{sdc}
S. W. Morris, E. Bodenschatz, D. S. Cannell, and G. Ahlers, Phys. Rev. 
Lett. {\bf 71}, 2026 (1993). 
\bibitem{liquidcrystal}
R. Ribotta and A. Joets, in {\it Cellular Structures and Instabilities},
edited by J. E. Wesfried and S. Zaslavski (Springer-Verlag, Berlin, 1984);
I. Rehberg, S. Rasenat, and V. Steinberg, Phys. Rev. Lett. {\bf 62}, 756 (1989).
\bibitem{chemical}
Q. Ouyang and H. L. Swinney, Chaos {\bf 1}, 411 (1991); Q. Ouyang, H. L.
Swinney, and G. Li, Phys. Rev. Lett. {\bf 84}, 1047 (2000).
\bibitem{faraday1}
A. B. Ezerski\u{i}, M. I. Rabinovich, V. P. Reutov, and I. M. 
Starobinets, Sov. Phys. JETP {\bf 64}, 1228 (1986),
A. Kudrolli and J. P. Gollub, Physica D, {\bf 97}, 133 (1996).
\bibitem{bretherton}
C. S. Bretherton and E. A. Spiegel, Phys. Lett. {\bf 96A}, 152 (1983).
\bibitem{shraiman}
B. I. Shraiman, Phys. Rev. Lett. {\bf 57}, 325 (1986), and references therein.
\bibitem{coullet89}
P. Coullet, L. Gil, and F. Rocca, Opt. Commun. {\bf 73}, 403 (1989).
\bibitem{eckmann91}
J. -P. Eckmann, and I. Procaccia, Phys. Rev. Lett. {\bf 66}, 891 (1991).
\bibitem{cross95}
M. C. Cross and Y. Tu, Phys. Rev. Lett. {\bf 75}, 834 (1995),
M. Cross, Physica D, {\bf 97}, 65 (1996).
\bibitem{thispattern1}
F. Melo, P. B. Umbanhowar, and H. L. Swinney, Phys. Rev. Lett. {\bf 72}, 
172 (1994).
\bibitem{bizon}
C. Bizon, M. D. Shattuck, J. B. Swift, W. D. McCormick, and H. L.
Swinney, Phys. Rev. Lett. {\bf 80}, 57 (1998).
\bibitem{walton}
O. R. Walton, in {\it Particulate Two-Phase Flow}, edited by M. C. Roco
(Butterworth-Heinemann, Boston, 1993), p. 884.
\bibitem{thispattern2}
F. Melo, P. B. Umbanhowar, and H. L. Swinney, Phys. Rev. Lett.  {\bf 75}, 
3838 (1995).
\bibitem{singleball}
Modeling an oscillated granular layer as a perfectly inelastic mass
is an old idea. Its dynamics with varying forcing was first systematically
studied in:
A. Mehta and J. M. Luck, Phys. Rev. Lett. {\bf 65}, 393 (1990).
\bibitem{aranson}
I. S. Aranson, D. Blair, W. K. Kwok, G. Karapetrov, U. Welp, G. W. Crabtree,
V. M. Vinokur, and L. S. Tsimring, Phys. Rev. Lett. {\bf 82}, 731 (1999).
\bibitem{douady}
S. Douady, S. Fauve, and C. Laroche, Europhys. Lett. {\bf 8}, 621 (1989).
\bibitem{wassgren}
C. R. Wassgren, C. E. Brennen, and M. L. Hunt, Trans. of the 
ASME {\bf 63}, 712 (1996).
\bibitem{lan}
Y. Lan and A. D. Rosato, Phys. Fluids {\bf 9}, 3615 (1997).

\end{references}
\end{document}